\documentclass[twocolumn]{aastex62}
\usepackage{graphicx, romannum, float, afterpage, amsmath, commath}

\begin{document}
\title{The Infrared Medium-deep Survey. \Romannum{7}. Faint Quasars at $z \sim 5$ in the ELAIS-N1 Field}
\shortauthors{Shin et al.}
\shorttitle{Faint Quasars at $z \sim 5$ in the ELAIS-N1 Field}

\correspondingauthor{Myungshin Im}
\email{myungshin.im@gmail.com}

\author[0000-0002-2188-4832]{Suhyun Shin}
\affil{Center for the Exploration of the Origin of the Universe (CEOU), Building 45, Seoul National University, 1 Gwanak-ro, Gwanak-gu, Seoul 08826, Republic of Korea}
\affil{Astronomy Program, FPRD, Department of Physics \& Astronomy, Seoul National University, 1 Gwanak-ro, Gwanak-gu, Seoul 08826, Republic of Korea}

\author[0000-0002-8537-6714]{Myungshin Im}
\affil{Center for the Exploration of the Origin of the Universe (CEOU), Building 45, Seoul National University, 1 Gwanak-ro, Gwanak-gu, Seoul 08826, Republic of Korea}
\affil{Astronomy Program, FPRD, Department of Physics \& Astronomy, Seoul National University, 1 Gwanak-ro, Gwanak-gu, Seoul 08826, Republic of Korea}

\author[0000-0003-1647-3286]{Yongjung Kim}
\affil{Kavli Institute for Astronomy and Astrophysics, Peking University, Beijing 100871, People’s Republic of China}
\affil{Center for the Exploration of the Origin of the Universe (CEOU), Building 45, Seoul National University, 1 Gwanak-ro, Gwanak-gu, Seoul 08826, Republic of Korea}
\affil{Astronomy Program, FPRD, Department of Physics \& Astronomy, Seoul National University, 1 Gwanak-ro, Gwanak-gu, Seoul 08826, Republic of Korea}

\author{Minhee Hyun}
\affil{Center for the Exploration of the Origin of the Universe (CEOU), Building 45, Seoul National University, 1 Gwanak-ro, Gwanak-gu, Seoul 08826, Republic of Korea}
\affil{Astronomy Program, FPRD, Department of Physics \& Astronomy, Seoul National University, 1 Gwanak-ro, Gwanak-gu, Seoul 08826, Republic of Korea}

\author[0000-0002-2548-238X]{Soojong Pak}
\affil{Center for the Exploration of the Origin of the Universe (CEOU), Building 45, Seoul National University, 1 Gwanak-ro, Gwanak-gu, Seoul 08826, Republic of Korea}
\affil{School of Space Research, Kyung Hee University, 1732 Deogyeong-daero, Giheung-gu, Yongin-si, Gyeonggi-do 17104, Republic of Korea}

\author[0000-0003-4847-7492]{Yiseul Jeon}
\affil{FEROKA Inc., 401, Open Innovation Building, Seoul Biohub, 117-3 Hoegi-ro, Dongdaemun-gu, Seoul 02455, Republic of Korea}

\author{Tae-Geun Ji}
\affil{Center for the Exploration of the Origin of the Universe (CEOU), Building 45, Seoul National University, 1 Gwanak-ro, Gwanak-gu, Seoul 08826, Republic of Korea}
\affil{School of Space Research, Kyung Hee University, 1732 Deogyeong-daero, Giheung-gu, Yongin-si, Gyeonggi-do 17104, Republic of Korea}

\author{Hojae Ahn}
\affil{Center for the Exploration of the Origin of the Universe (CEOU), Building 45, Seoul National University, 1 Gwanak-ro, Gwanak-gu, Seoul 08826, Republic of Korea}
\affil{School of Space Research, Kyung Hee University, 1732 Deogyeong-daero, Giheung-gu, Yongin-si, Gyeonggi-do 17104, Republic of Korea}

\author{Seoyeon Byeon}
\affil{Center for the Exploration of the Origin of the Universe (CEOU), Building 45, Seoul National University, 1 Gwanak-ro, Gwanak-gu, Seoul 08826, Republic of Korea}
\affil{School of Space Research, Kyung Hee University, 1732 Deogyeong-daero, Giheung-gu, Yongin-si, Gyeonggi-do 17104, Republic of Korea}

\author{Jimin Han}
\affil{Center for the Exploration of the Origin of the Universe (CEOU), Building 45, Seoul National University, 1 Gwanak-ro, Gwanak-gu, Seoul 08826, Republic of Korea}
\affil{School of Space Research, Kyung Hee University, 1732 Deogyeong-daero, Giheung-gu, Yongin-si, Gyeonggi-do 17104, Republic of Korea}

\author{Sungyong Hwang}
\affil{Center for the Exploration of the Origin of the Universe (CEOU), Building 45, Seoul National University, 1 Gwanak-ro, Gwanak-gu, Seoul 08826, Republic of Korea}
\affil{Astronomy Program, FPRD, Department of Physics \& Astronomy, Seoul National University, 1 Gwanak-ro, Gwanak-gu, Seoul 08826, Republic of Korea}

\author{Sophia Kim}
\affil{Center for the Exploration of the Origin of the Universe (CEOU), Building 45, Seoul National University, 1 Gwanak-ro, Gwanak-gu, Seoul 08826, Republic of Korea}
\affil{Astronomy Program, FPRD, Department of Physics \& Astronomy, Seoul National University, 1 Gwanak-ro, Gwanak-gu, Seoul 08826, Republic of Korea}

\author{Gu Lim}
\affil{Center for the Exploration of the Origin of the Universe (CEOU), Building 45, Seoul National University, 1 Gwanak-ro, Gwanak-gu, Seoul 08826, Republic of Korea}
\affil{Astronomy Program, FPRD, Department of Physics \& Astronomy, Seoul National University, 1 Gwanak-ro, Gwanak-gu, Seoul 08826, Republic of Korea}

\author{Insu Paek}
\affil{Center for the Exploration of the Origin of the Universe (CEOU), Building 45, Seoul National University, 1 Gwanak-ro, Gwanak-gu, Seoul 08826, Republic of Korea}
\affil{Astronomy Program, FPRD, Department of Physics \& Astronomy, Seoul National University, 1 Gwanak-ro, Gwanak-gu, Seoul 08826, Republic of Korea}

\author{Gregory S. H. Paek}
\affil{Center for the Exploration of the Origin of the Universe (CEOU), Building 45, Seoul National University, 1 Gwanak-ro, Gwanak-gu, Seoul 08826, Republic of Korea}
\affil{Astronomy Program, FPRD, Department of Physics \& Astronomy, Seoul National University, 1 Gwanak-ro, Gwanak-gu, Seoul 08826, Republic of Korea}

\author{Yoon Chan Taak}
\affil{Center for the Exploration of the Origin of the Universe (CEOU), Building 45, Seoul National University, 1 Gwanak-ro, Gwanak-gu, Seoul 08826, Republic of Korea}
\affil{Astronomy Program, FPRD, Department of Physics \& Astronomy, Seoul National University, 1 Gwanak-ro, Gwanak-gu, Seoul 08826, Republic of Korea}

\author{Changsu Choi}
\affil{Center for the Exploration of the Origin of the Universe (CEOU), Building 45, Seoul National University, 1 Gwanak-ro, Gwanak-gu, Seoul 08826, Republic of Korea}
\affil{Astronomy Program, FPRD, Department of Physics \& Astronomy, Seoul National University, 1 Gwanak-ro, Gwanak-gu, Seoul 08826, Republic of Korea}

\author{Jueun Hong}
\affil{Center for the Exploration of the Origin of the Universe (CEOU), Building 45, Seoul National University, 1 Gwanak-ro, Gwanak-gu, Seoul 08826, Republic of Korea}
\affil{Astronomy Program, FPRD, Department of Physics \& Astronomy, Seoul National University, 1 Gwanak-ro, Gwanak-gu, Seoul 08826, Republic of Korea}

\author[0000-0003-1470-5901]{Hyunsung David Jun}
\affil{Korea Institute for Advanced Study (KIAS), 85 Hoegiro, Dongdaemun-gu, Seoul, 02455, Republic of Korea}

\author[0000-0002-6925-4821]{Dohyeong Kim}
\affil{Kavli Institute for Astronomy and Astrophysics, Peking University, Beijing 100871, People’s Republic of China}
\affil{Center for the Exploration of the Origin of the Universe (CEOU), Building 45, Seoul National University, 1 Gwanak-ro, Gwanak-gu, Seoul 08826, Republic of Korea}
\affil{Astronomy Program, FPRD, Department of Physics \& Astronomy, Seoul National University, 1 Gwanak-ro, Gwanak-gu, Seoul 08826, Republic of Korea}

\author[0000-0001-5120-0158]{Duho Kim}
\affil{Arizona State University, School of Earth and Space Exploration, P.O. Box 871404, Tempe, AZ 85287-1404, USA}
\affil{Korea Astronomy and Space Science Institute, 776, Daedeokdae-ro, Yuseong-gu, Daejeon 34055, Republic of Korea}

\author[0000-0002-3560-0781]{Minjin Kim}
\affil{Department of Astronomy and Atmospheric Sciences, College of Natural Sciences, Kyungpook National University, Daegu 41566, Republic of Korea}

\author[0000-0002-1710-4442]{Jae-Woo Kim}
\affil{Korea Astronomy and Space Science Institute, 776, Daedeokdae-ro, Yuseong-gu, Daejeon 34055, Republic of Korea}

\author[0000-0002-1418-3309]{Ji Hoon Kim}
\affil{Subaru Telescope, National Astronomical Observatory of Japan, 650 North A’ohoku Place, Hilo, HI 96720, USA}
\affil{METASPACE Inc., 401  36, Nonhyeon-ro, Gangnam-gu, Seoul, Republic of Korea}

\author{Hye-In Lee}
\affil{Center for the Exploration of the Origin of the Universe (CEOU), Building 45, Seoul National University, 1 Gwanak-ro, Gwanak-gu, Seoul 08826, Republic of Korea}
\affil{School of Space Research, Kyung Hee University, 1732 Deogyeong-daero, Giheung-gu, Yongin-si, Gyeonggi-do 17104, Republic of Korea}

\author[0000-0001-5342-8906]{Seong-Kook Lee}
\affil{Center for the Exploration of the Origin of the Universe (CEOU), Building 45, Seoul National University, 1 Gwanak-ro, Gwanak-gu, Seoul 08826, Republic of Korea}
\affil{Astronomy Program, FPRD, Department of Physics \& Astronomy, Seoul National University, 1 Gwanak-ro, Gwanak-gu, Seoul 08826, Republic of Korea}

\author{Won-Kee Park}
\affil{Korea Astronomy and Space Science Institute, 776, Daedeokdae-ro, Yuseong-gu, Daejeon 34055, Republic of Korea}

\author[0000-0001-8012-5871]{Woojin Park}
\affil{Center for the Exploration of the Origin of the Universe (CEOU), Building 45, Seoul National University, 1 Gwanak-ro, Gwanak-gu, Seoul 08826, Republic of Korea}
\affil{School of Space Research, Kyung Hee University, 1732 Deogyeong-daero, Giheung-gu, Yongin-si, Gyeonggi-do 17104, Republic of Korea}

\author{Yongmin Yoon}
\affil{Korea Institute for Advanced Study (KIAS), 85 Hoegiro, Dongdaemun-gu, Seoul, 02455, Republic of Korea}
\affil{Center for the Exploration of the Origin of the Universe (CEOU), Building 45, Seoul National University, 1 Gwanak-ro, Gwanak-gu, Seoul 08826, Republic of Korea}
\affil{Astronomy Program, FPRD, Department of Physics \& Astronomy, Seoul National University, 1 Gwanak-ro, Gwanak-gu, Seoul 08826, Republic of Korea}

\begin{abstract}
The intergalactic medium (IGM) at $z\sim$ 5 to 6 is largely ionized, and yet the main source for the IGM ionization in the early universe is uncertain. Of the possible contributors are faint quasars with $-26 \lesssim M_{\rm 1450} \lesssim -23$, but their number density is poorly constrained at $z\sim5$. In this paper, we present our survey of faint quasars at $z\sim5$ in the European Large-Area {\it ISO} Survey-North 1 (ELAIS-N1) field over a survey area of 6.51 deg$^2$ and examine if such quasars can be the dominant source of the IGM ionization. We use the deep optical/near-infrared data of the ELAIS-N1 field as well as the additional medium-band observations to find $z \sim 5$ quasars through a two-step approach using the broadband color selection, and SED fitting with the medium-band information included. Adopting Bayesian information criterion, we identify ten promising quasar candidates. Spectra of three of the candidates are obtained, confirming all of them to be quasars at $z\sim5$ and supporting the reliability of the quasar selection. Using the promising candidates, we derive the $z\sim5$ quasar luminosity function at $-26 \lesssim M_{\rm 1450} \lesssim -23$. The number density of faint $z\sim5$ quasars in the ELAIS-N1 field is consistent with several previous results that quasars are not the main contributors to the IGM-ionizing photons at $z\sim5$.
\end{abstract}

\section{Introduction} \label{sec:intro}
\indent One of the most pivotal cosmic events after the Big Bang is cosmic reionization. Cosmic reionization is the change in the ionization state of the intergalactic medium (IGM), where the neutral atoms in the IGM become ionized after the recombination era due to ultra-violet (UV) radiation from newly born stars, galaxies, and active galactic nuclei (AGNs). Hence, the cosmic reionization has become an important subject of study for understanding the history of astronomical objects in the universe. The recent analysis of the cosmic microwave background suggests that mid-point of reionization is at $z\sim7.7$ (Planck Collaboration 2018), corresponding to the most distant quasars discovered so far (Ba{\~n}ados et al. 2018; Davies et al. 2018). The observational constraints from high-redshift observation indicate the cosmic reionization ends by $z\sim6$ (Bouwens et al. 2015).

\indent However, the nature of the sources responsible for keeping the IGM ionized in the post-reionization era remains uncertain. Previous studies have suggested either star-forming galaxies or quasars as candidates to explain the majority of the required photon budget (Fontanot et al. 2012; Giallongo et al. 2015, hereafter G15; Madau \& Haardt 2015; Bouwens et al. 2016; D'Aloisio et al. 2017; Ricci et al. 2017; Kakiichi et al. 2018). In some works, galaxies are considered as the primary contributors to IGM reionization (Alvarez et al. 2012; Paardekooper et al. 2013; Kashikawa et al. 2015; Kim et al. 2015, 2019; Matsuoka et al. 2018; Smith et al. 2018; Fletcher et al. 2019). In contrary, the escape fraction of the hydrogen-ionizing Lyman-continuum (LyC) photons of star-forming galaxies is found to be only few percent (Bridge et al. 2010; Rutkowski et al. 2016; Vasei et al. 2016; Japelj et al. 2017; Iwata et al. 2019), and this makes it challenging for galaxies to maintain the ionized state of the IGM (Finkelstein et al. 2015; Grazian et al. 2017; Naidu et al. 2018; Steidel et al. 2018; Lam et al. 2019). 

\indent On the other hand, the escape fraction of LyC photons in quasars is about 100\% (Cristiani et al. 2016; Guaita et al. 2016; Grazian et al. 2018; Romano et al. 2019), and hence several studies suggest that quasars are the main IGM ionizing source (Feng et al. 2016; G15; Grazian et al. 2016; Boutsia et al. 2018; Romano et al. 2019). Given the spectral shape and the escape fraction of LyC photons of quasars, the contribution of quasars to the hydrogen reionizing photons can be calculated from the quasar luminosity function (LF). Therefore, many observational studies have tried to determine the LF of quasars at $z > 4$ (Barger et al. 2003; Fan et al. 2006; Jiang et al. 2008; Willott et al. 2010; McGreer et al. 2013; Ba\~naods et al. 2016; Yang et al. 2016; Jeon et al. 2017; Kim et al. 2019). 

\indent Although still uncertain, the quasar LFs show that the number of LyC photons is the most sensitive to the number of quasars with moderate luminosity at $-24 < M_{1450} < -23$. Interestingly, some studies (G15, Giallongo et al. 2019, hereafter G19; Boutsia et al. 2018) find that the number density of the moderate luminosity quasars is higher by a factor of $\sim$ 10 than other studies (Onoue et al. 2017; Akiyama et al. 2018; Matsuoka et al. 2018; McGreer et al. 2018, hereafter M18). The high number density in the moderate luminosity range enables quasars to be the main sources for the IGM ionization. The conflict in the high redshift quasar luminosity function reflects the lack of quasar samples with $M_{\rm 1450} < -24$. It is, therefore, imperative to add more samples to help settle the issue. 

\indent In order to constrain the quasar LF better and understand the IGM ionization process, we have carried out the Infrared Medium-deep Survey (IMS; M. Im et al. 2020, in preparation), a deep and wide near-infrared (NIR) survey to depths of $J \sim 23.5$ AB mag and with an area coverage of $\sim$ 100 deg$^2$. Color selection criteria with the deep NIR data can discern reliable high redshift quasars from dwarf stars, the most contaminant source for quasar candidate selection (McGreer et al. 2013). Additionally, the medium-band imaging of this survey improves the spectral energy distribution (SED) fittings of a quasar and a M-dwarf star. Adopting the Bayesian information criterion (BIC) which can consider the number of free parameters in a model, we are able to compare the two models and select the promising quasar candidates effectively. In this paper, we focus on our discovery of faint quasars at $z \sim 5$ in the European Large-Area {\it ISO} Survey-North 1 (ELAIS-N1) field. The extensive multi-wavelength coverage of the ELAIS-N1 field, including the NIR from IMS, the optical from the Hyper Suprime-Cam Subaru Strategic Program (HSC-SSP) and medium-bands, makes it possible to select faint $z \sim 5$ quasars effectively.

\indent This paper is structured as the following. Section 2 describes the SED models for quasars and M dwarfs. In Section 3, we present the broadband and medium-band imaging data used for the quasar selection. In Section 4, we explain two quasar selection methods, the selection criteria based on broadband colors and BIC selection based on the SED-fitting. Follow-up spectroscopy of quasar candidates using the MMT is presented in Section 5. In Section 6, we build the quasar LF and discuss the role of quasars in the IGM reionization. Finally, we summarize our findings. Throughout this paper, we adopt the AB magnitude system for all filters (Oke \& Gunn 1983) and assume $H_{0} = 70$ km s$^{-1}$ Mpc$^{-1}$, $\Omega_{M} = 0.3$, and $\Omega_{\Lambda} = 0.7$ of the concordance $\Lambda$CDM cosmology, which has been supported by the observational studies in the past decades (e.g., Im et al. 1997; Planck Collaboration 2018).

\section{SED model} \label{sec:style}
In this study, two methods based on SED models are used to select quasar candidates: broadband color cuts (Section 4.1) and BIC (Section 4.2). Below, we describe the quasar and M dwarf star SED models we used for defining the quasar selection criteria.

\subsection{Quasar SED model}
To establish selection criteria, we create various quasar SED models by adopting the composite quasar spectrum of Vanden Berk et al. (2001) as a base model and varying the continuum slope ($\alpha_{\lambda}$) and the equivalent width (EW) of Ly$\alpha$ and N \Romannum{5} $\lambda 1240$. The Selsing et al. (2016) quasar SED template may be useful in that it reduces the host galaxy contribution near 5000 \AA{} in rest frame, but they construct the SED template based on very luminous blue quasars with $M_{i} \sim -29$ at $z\sim 1-2$  which is inadequate to find faint quasars with $M_{\rm 1450} \sim -23.5$. Regarding Brown et al. (2019), they make use of the SED spanning UV to radio based on only 41 AGNs. On the other hand, Vanden Berk et al. (2001) generate the SED template using over 2,000 spectra of quasars at $0<z<5$. Thus, we use Vanden Berk et al. (2001) composite spectra to generate diverse quasar SED models.

\indent To generate realistic quasar models, we take the empirical distribution of $\alpha_{\lambda}$ with $\langle \alpha_{\lambda}\rangle = -1.60$ and $\sigma (\alpha_{\lambda}) = 1.0$ (Mazzucchelli et al. 2017) where $F_{\lambda} \propto  \lambda^{\alpha_{\lambda}}$.  This relation is obtained from 15 quasars at $z \gtrsim 6.5$, but the mean continuum slope is comparable to $-1.54$ of Vanden Berk et al. (2001) SED and $-1.70$ of Selsing et al. (2016) SED. From the above values, the mean continuum slope of quasars might have consistent values across the luminosity and redshift range. So, we adjust a given continuum slope $\alpha_{\lambda}$ by multiplying a factor of $(\lambda / 1000$ \AA)$^{\alpha_{\lambda} - \alpha_{\lambda,\rm{mean}}}$ where $\alpha_{\lambda, \rm{mean}} = -1.6$.

\indent Similarly, we use the EW distribution of Ly$\alpha$+N\Romannum{5} line from Ba\~nados et al. (2016) which has $\langle$log EW (\AA)$\rangle = 1.542$ and $\sigma$(log EW (\AA))$= 0.391$. To calculate the EW, we model a local continuum slope at a wavelength range of 1216--1800 \AA, excluding wavelength ranges corresponding to several emission lines. Then, we integrate the fluxes over the continuum model at a rest frame wavelength range of 1160--1290 \AA. To make a SED with a given EW, we replace the continuum-subtracted flux with EW$_{0} = 92.91$ ($f_{\rm{EW},0}$) to the rescaled value ($f_{\rm{EW}}$).

\indent To describe the IGM absorption, we adopt the IGM attenuation model of Inoue et al. (2014; hereafter, I14). Note that the I14 model's correction to the rest-frame wavelength of 912--1216 \AA{} is somewhat less than the popular IGM attenuation model of Madau (1995), in a sense that the observed $g$ and $r$ magnitudes of $z \sim 5$ model SEDs become brighter.

\indent We set the redshift range of 4.5 to 5.6 with a grid size of 0.01, the $\alpha_{\lambda}$ range of -3.6 to 0.2 with a grid size of 0.2, the logarithmic value of EW range of 0.5 to 2.5 with a grid size of 0.2, and the absolute magnitude at 1450 \AA{} in the rest frame ($M_{\rm 1450}$) range of $-27 < M_{1450} < -21$ with a grid size of 0.1 mag, resulting in $\sim 10^7$ SEDs in total.

\subsection{M dwarf model}
When selecting quasars from colors and SED shapes, the main contaminants are low mass stars, in particular, M-dwarf stars. Therefore, we use the M-dwarf spectra from the BT-settl model (Allard et al. 2013). The spectra are known to match the observed M dwarf spectra on a wide range of physical parameters. The spectra are available from a public archive {Theoretical Spectra Web Server}\footnote{\url{http://svo2.cab.inta-csic.es/theory/newov/index.php}}. Considering typical ranges of M dwarfs' parameters (Casagrande et al. 2008; Rajpurohit et al. 2013), we use a total of 1,050 spectra covering the parameter space of $T_{\text{eff}}$ of 2000--4000 K with a 100 K step, log($g$) of 3.5--6.0 with a 0.5 step, [M/H] of -4--0.5 with 0.2 $\sim$ 0.5 steps, and [$\alpha$/M] of $-0.2$--0.4 with a 0.2 step. We add a normalization factor (f$_{N}$) as a free parameter.

\section{Imaging Data}

\begin{figure}[t!]
\centering
\includegraphics[scale=0.28]{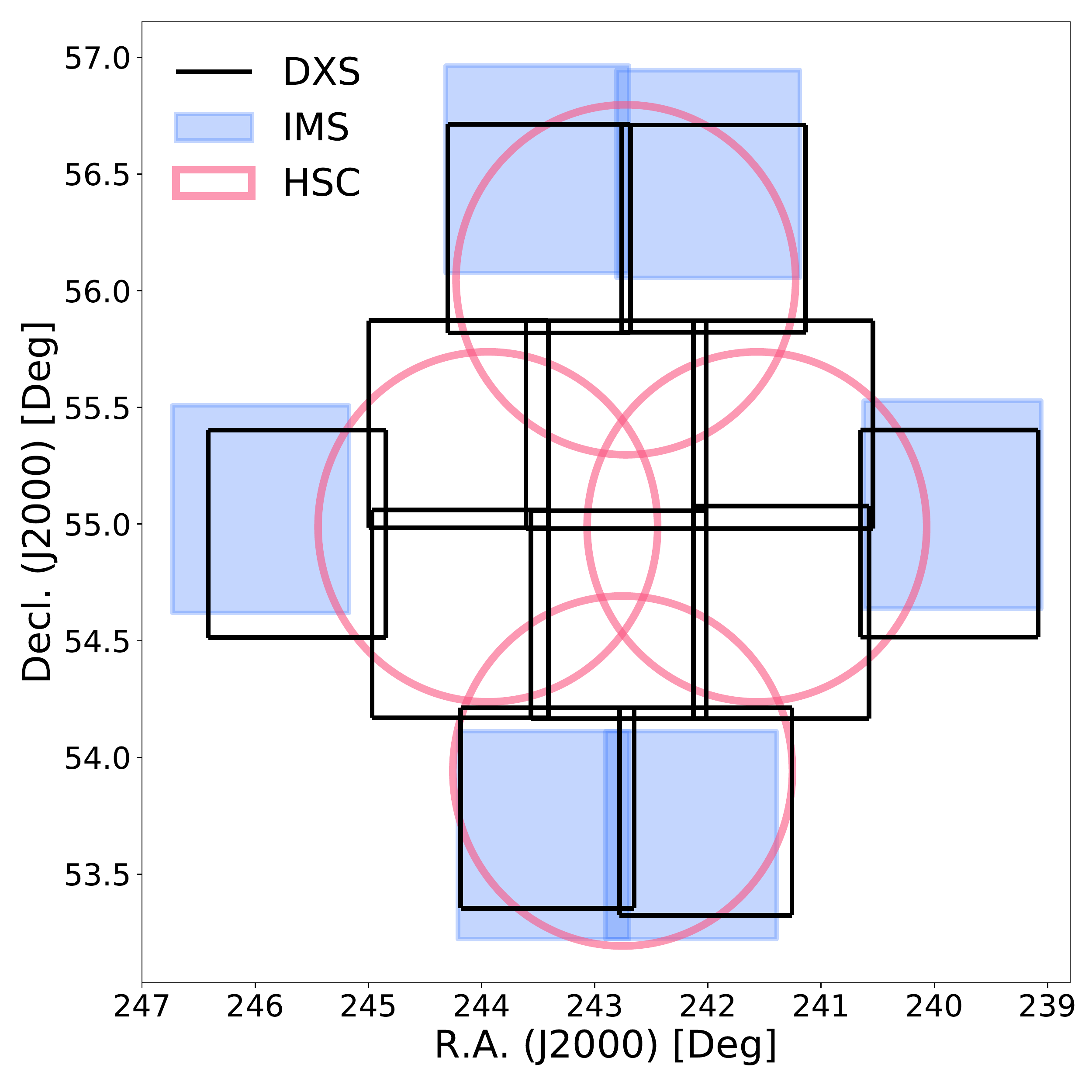}
\caption{ELAIS-N1 area coverages of various surveys. The black rectangles show the DXS coverage. The IMS data provides additional $J$-band data in the outer part of the ELAIS-N1 field (filled blue rectangles). The HSC optical imaging data are plotted in pink circles.}
\label{fig:field}
\end{figure}

\begin{figure*}[tbh] 
\centering
\includegraphics[scale=0.24]{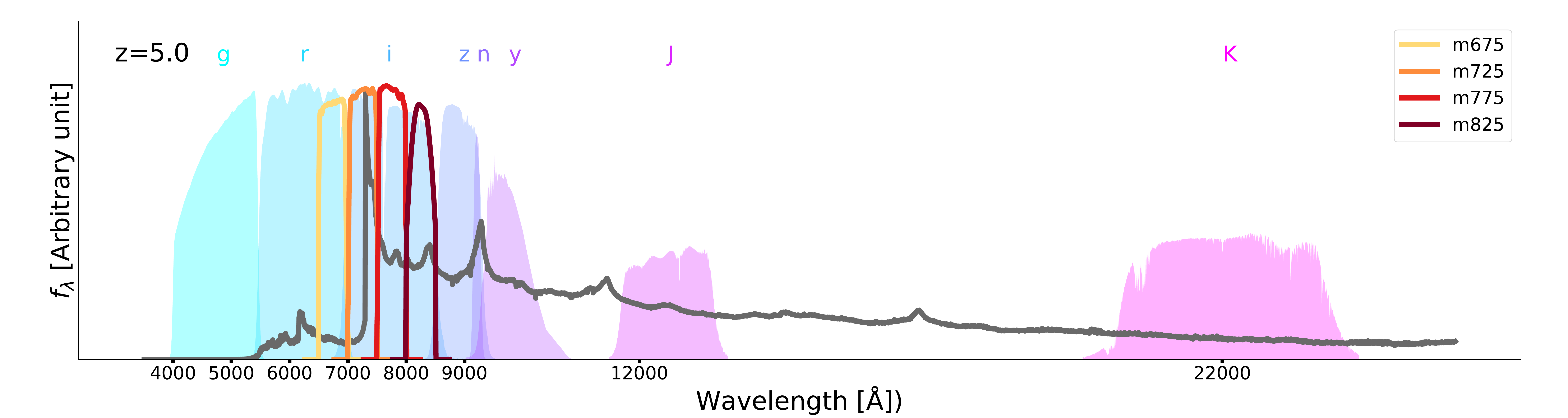}
\caption{Filter transmission curves (including quantum efficiency) of the data we used in this work. For the sake of easy distinction, the filter curves of HSC-SSP ($g,r,i,z,NB921$ and $y$) and DXS/IMS ($J$, and $K$) are drawn in background, while the medium-band filters are plotted as the solid lines. The dark gray line shows the representative SED of a quasar at $z=5.0$.}
\label{fig:filtersystem}
\end{figure*}

\subsection{broadband data}
The ELAIS-N1 field is one of the deep extragalactic survey fields observed by the Infrared Space Observatory (ISO) due to the low infrared background (Rowan-Robinson et al. 2004; Vaccari et al. 2005). The ELAIS-N1 field is centered at 16:11:00 +55:00:00 (J2000), and a wealth of deep, wide-area, and multi-wavelength data are available for this field. Our quasar selection requires deep optical and NIR imaging data. Therefore, HSC-SSP (DR1, Aihara et al. 2018a, 2018b) is chosen for optical data and the 3.8m United Kingdom Infrared Telescope (UKIRT) Infrared Deep Sky Survey (UKIDSS) – Deep Extragalactic Survey (DXS; Lawrence et al. 2007) or the Infrared Medium-deep Survey (IMS) are chosen for NIR data. The field coverages of these surveys are plotted in Figure~\ref{fig:field} according to the instruments' field of view and pointing. Figure~\ref{fig:filtersystem} shows all the filters from these survey that are used in this study.

\subsubsection{Optical -- HSC}
For the optical data, we retrieved a catalog from the HSC data archive system. HSC-SSP is a three-layered imaging survey with each layer survey covering 1400 deg$^2$ (Wide), 27 deg $^2$ (Deep), 5 deg $^2$ (Ultra-deep) with the target image depths of $r \sim 26.1$, 27.1, and 27.7 mags respectively for 5-$\sigma$, point source detection (Aihara et al. 2018a). The ELAIS-N1 field is one of the HSC-SSP deep fields of which DR1 $i$-band depth reaches $\sim$ 26.5 mag (Aihara et al. 2018b). Following the method in Matsuoka et al. 2018, which used a random source catalog from HSC-SSP archival system, we found the effective HSC-SSP DR1 coverage of ELAIS-N1 to be 6.75 deg$^2$. Images of this field in S17A season were taken with multiple filters including $g,r,i,z,y$, and a narrow-band centered at $9210$ \AA{} ($NB921$), reaching 5-$\sigma$ depths of 26.8, 26.6, 26.5, 25.6, 24.8, and 25.6 mag, respectively, for point-sources (Aihara et al. 2018b). Considering that Ly$\alpha$ break is expected to be at $\sim 7200 \rm{\AA}$ at $z\sim5$, we used $i$-band detected catalog from HSC-SSP database (Aihara et al. 2018a; Aihara et al. 2018b). A quasar would appear as a point source under the seeing condition of HSC-SSP images. Therefore, we used the point-spread function (PSF) magnitudes in the HSC-SSP catalog for the optical photometry. Note that all magnitudes in the catalog are extinction-corrected values according to the dust map of the Schlegel et al. (1998, hereafter SFD).

\subsubsection{NIR -- DXS/IMS}
DXS and IMS images were obtained with WFCAM at UKIRT. The combination of the DXS and IMS data covers almost the entire HSC ELAIS-N1 area in $J$-band. On the other hand, $K$-band image is only available in an area covered by DXS. The 5 $\sigma$ imaging depths are $J \sim 23.2$ and $K \sim 22.7$ mag for point sources (Hewett et al. 2006; M. Im et al. 2020, in preparation). Note that the $J$-band depths of the IMS and DXS are nearly identical and the total NIR survey area is 9.68 deg$^{2}$. We performed the photometry of IMS and DXS data with our own pipeline that uses SExtractor (Bertin $\&$ Arnouts 1996). The auto-magnitude were used as total magnitudes. We found that the magnitudes of bright stars are consistent with those of stars in the Two Micron All Sky Survey (2MASS; Skrutskie et al. 2006). Extinction correction values of these bands were calculated using the wavelength-dependent relation (Cardelli et al. 1989) between $A(V)$ and $E(B-V)$ from SFD assuming $R_{V}$ = 3.1 and a representative wavelength equal to their effective wavelength. 

\begin{deluxetable*}{cccccccc} 
\tablenum{1}
\tablecaption{Photometry of our $z\sim5$ quasar candidates} 
\tablehead{\colhead{ID} & \colhead{$g$} & \colhead{$r$} & \colhead{$i$} & \colhead{$NB921$} & \colhead{$z$} & \colhead{$J$} & \colhead{$K$} \\ \colhead{} & \colhead{(mag)} & \colhead{(mag)} & \colhead{(mag)} & \colhead{(mag)} & \colhead{(mag)} & \colhead{(mag)} }
\startdata
IMS J160306+541928 & $>$ 26.8 & 24.88 $\pm$ 0.17 & 23.13 $\pm$ 0.03 & 22.41 $\pm$ 0.02 & 22.71 $\pm$ 0.21 & 22.73 $\pm$ 0.22 & 22.99 $\pm$ 0.30 \\
IMS J160517+554002 & $>$ 26.8 & 24.31 $\pm$ 0.05 & 22.57 $\pm$ 0.01 & 22.46 $\pm$ 0.01 & 22.31 $\pm$ 0.02 & 21.92 $\pm$ 0.12 & 21.78 $\pm$ 0.14 \\
IMS J160552+555340 & $>$ 26.8 & 23.31 $\pm$ 0.02 & 21.45 $\pm$ 0.00 & 21.08 $\pm$ 0.01 & 21.00 $\pm$ 0.01 & 21.27 $\pm$ 0.07 & 20.54 $\pm$ 0.06 \\
IMS J160622+540056 & $>$ 26.8& 26.19 $\pm$ 0.21 & 23.49 $\pm$ 0.01 & 23.40 $\pm$ 0.04 & 23.24 $\pm$ 0.04 & 22.70 $\pm$ 0.17 & $>$ 22.700 \\
IMS J160732+544750 & $>$ 26.8 & 24.68 $\pm$ 0.05 & 23.44 $\pm$ 0.01 & 23.20 $\pm$ 0.02 & 23.21 $\pm$ 0.03 & $>$ 23.158 & $>$ 22.700 \\
IMS J160748+541157 & $>$ 26.8 & 24.45 $\pm$ 0.04 & 22.90 $\pm$ 0.01 & 22.65 $\pm$ 0.02 & 22.56 $\pm$ 0.02 & 22.49 $\pm$ 0.15 & 21.64 $\pm$ 0.15 \\
IMS J160914+554511 & $>$ 26.8 & 24.06 $\pm$ 0.03 & 22.55 $\pm$ 0.01 & 22.42 $\pm$ 0.01 & 22.31 $\pm$ 0.01 & 22.82 $\pm$ 0.18 & 22.85 $\pm$ 0.26 \\
IMS J161248+550927 & $>$ 26.8 & 24.19 $\pm$ 0.03 & 22.78 $\pm$ 0.01 & 22.62 $\pm$ 0.02 & 22.52 $\pm$ 0.02 & 22.67 $\pm$ 0.18 & 22.36 $\pm$ 0.20 \\
IMS J161341+542146 & $>$ 26.8 & 24.94 $\pm$ 0.05 & 23.49 $\pm$ 0.01 & 23.29 $\pm$ 0.02 & 23.29 $\pm$ 0.03 & 23.31 $\pm$ 0.25 & 0.00 $\pm$ 0.00 \\
IMS J161343+542131 & 25.24 $\pm$ 0.03 & 22.94 $\pm$ 0.01 & 20.58 $\pm$ 0.00 & 19.80 $\pm$ 0.00 & 19.92 $\pm$ 0.00 & 19.40 $\pm$ 0.02 & 19.61 $\pm$ 0.03 \\
IMS J161636+535545 & $>$ 26.8 & 24.86 $\pm$ 0.08 & 23.23 $\pm$ 0.02 & 22.70 $\pm$ 0.03 & 22.85 $\pm$ 0.05 & $>$ 23.158 & $>$ 22.700 \\
IMS J161827+551748 & 25.39 $\pm$ 0.04 & 22.78 $\pm$ 0.01 & 21.11 $\pm$ 0.00 & 20.87 $\pm$ 0.00 & 20.83 $\pm$ 0.00 & 20.78 $\pm$ 0.05 & 20.14 $\pm$ 0.04 \\
IMS J161903+545638 & 25.78 $\pm$ 0.05 & 23.95 $\pm$ 0.02 & 22.39 $\pm$ 0.01 & 21.98 $\pm$ 0.01 & 22.06 $\pm$ 0.01 & 21.78 $\pm$ 0.10 & 22.11 $\pm$ 0.20 \\
\enddata
\tablecomments{According to Table 2 in Aihara et al. (2018b) the 5-$\sigma$ depth in the $g$-band is 26.8 mag for a point-source.}
\label{table:grizJmag}
\end{deluxetable*}

\subsubsection{Catalog matching}
The optical and NIR catalogs were matched with a matching radius of $1\farcs0$. When an object was not detected in $J$ or $K$, we assigned $J$ or $K$ limiting magnitudes to the object. If the matching resulted in more than one matched pair, we chose an object in the NIR catalog that is closer to the optical source or has magnitudes similar to the optical data. Multiple matching case occurred in $\sim 2~\%$ of all the matched objects.

\indent Our quasar search was performed on the overlap area of the optical and the NIR data, and we calculated this quasar search area in the following way. The NIR sources were matched with the non-saturated ($i > 19$) and bright ($i < 21$) optical sources in the HSC catalog with a matching distance of $1\farcs0$. The bright magnitude cut at $i < 21$ ensures that each $i$-band source must have a NIR counterpart in the $J$-band catalog in the area where the optical and NIR images overlap. Then, the fraction of the optical sources with NIR counterparts should be identical to the fraction of the optical image area that overlaps with the NIR image. The 96.4 $\%$ of the optical sources were found to have NIR counterparts, thus giving 6.51 deg$^{2}$ as the quasar search area.

\begin{deluxetable*}{ccccccccccccccc}
\tablenum{2}
\tablecaption{Medium-band photometry of the broadband selected quasar candidates} 
\tablehead{\colhead{ID} & \colhead{R.A.} & \colhead{Decl.} & \colhead{$m675$} & \colhead{$m725$} & \colhead{$m775$} & \colhead{$m825$} \\ \colhead{} & \colhead{(J2000)} & \colhead{(J2000)} & \colhead{(mag)} & \colhead{(mag)} & \colhead{(mag)} & \colhead{(mag)} }
\startdata
IMS J160306+541928 & 16:03:06.02 & 54:19:28.46 & $>$ 24.17 & 22.98 $\pm$ 0.18 & &  \\
IMS J160517+554002 & 16:05:17.80 & 55:40:02.00 &  & $>$23.55 & 21.96 $\pm$ 0.12 & \\
IMS J160552+555340 & 16:05:52.07 & 55:53:40.61 &  & $>$23.06 & 20.87 $\pm$ 0.10 & \\
IMS J160622+540056 & 16:06:22.50 & 54:00:56.64 &  &  &  &  \\
IMS J160732+544750 & 16:07:32.21 & 54:47:50.86 & $>$24.19 & 22.25 $\pm$ 0.10 & & \\
IMS J160748+541157 & 16:07:48.15 & 54:11:57.40 & $>$24.29 & 22.90 $\pm$ 0.26 & & \\
IMS J160914+554511 & 16:09:14.68 & 55:45:11.83 & $>$23.23 & 21.94 $\pm$ 0.18 & 22.65 $\pm$ 0.14 &  \\
IMS J161248+550927 & 16:12:48.98 & 55:09:27.54 & $>$23.48 & 22.79 $\pm$ 0.17\tablenotemark{a} &  23.00 $\pm$ 0.20 &  \\
IMS J161341+542146 & 16:13:41.15 & 54:21:46.65 & & & & \\
IMS J161343+542131 & 16:13:43.43 & 54:21:31.29 & 21.44 $\pm$ 0.10 & 21.00 $\pm$ 0.06 & 20.81 $\pm$ 0.11 & \\
IMS J161636+535545 & 16:16:36.32 & 53:55:45.01 & & & & \\
IMS J161827+551748 & 16:18:27.29 & 55:17:48.47 & $>$23.09 & 21.31 $\pm$ 0.21 & 21.55 $\pm$ 0.21 & 21.34 $\pm$ 0.20 \\
IMS J161903+545638 & 16:19:03.74 & 54:56:38.92 &   & 22.18 $\pm$ 0.21 & 22.50 $\pm$ 0.18 &  & \\
\enddata
\tablenotetext{a}{Observed in Maidanak Observatory}
\label{table:mediummag}
\end{deluxetable*}

\subsection{Medium-band data}
Multiple steps were taken to select $z \sim 5$ quasar candidates, starting with a selection using the broadband data (Section 4.1). Then, we refined the candidate selection using the SED-fitting method (Section 4.2). To improve the SED-fitting, we obtained additional medium-band data of the broadband selected candidates. The medium-bands that we used have widths of $\sim$ 500 \AA{} centered at 675, 725, 775, and 825 nm (hereafter, $m675$, $m725$, $m775$, and $m825$). Compared to the broadbands with widths of $\sim 1500$ \AA{}, the medium-bands can improve the spectral sampling by a factor of 3, and hence the better quasar selection can be achieved. The medium-band observations were conducted using the SED camera for QUasars in EArly uNiverse (SQUEAN) on the 2.1-m Otto-Struve telescope at Mcdonald Observatory (Park et al. 2012; Kim et al. 2016) and Seoul National University 4K x 4K Camera (SNUCAM) on the 1.5-m telescope at Maidanak Observatory (Im et al. 2010). The data were taken with the seeing condition better than $1\farcs5$ with the on-source exposure time shorter than two hours, or two hours at maximum. The observations and targets are summarized in Table~\ref{table:mediummag}. In total, 10 out of 13 candidates have been observed.

\indent All the medium-band images were pre-processed including bias/dark subtraction and flat fielding. In Maidanak data, a fringe pattern appeared in the images, and they were corrected by subtracting the master fringe frame made from dithered observed images, following Jeon et al. (2010). Astrometry solutions were obtained for all the pre-processed images using the Astrometry.net software (Lang et al. 2010). 

\indent Then, the images were background-subtracted and flux-rescaled based on the zero-point of each image to fill the gap of various circumstances of observational conditions at each observational period. Outlier images determined by seeing and depth distribution of images were excluded in combining procedure.

\indent The zero-point of each medium-band image was derived by following the prescription described in Jeon et al. (2016) and Choi \& Im (2017). The procedure first identified bright but non-saturated stars near targets with existing multi-wavelength photometry data. In this case, we used $r, i$ and $z$ magnitudes of $5\sim10$ stars from the Seventh Data Release of the Sloan Digital Sky Survey (SDSS-DR7; Abazajian et al. 2009). Then, for each star, the best-fit stellar template was identified among 175 stellar templates of Gunn \& Stryker (1983). The medium-band photometry of the star was derived using the best-fit template, and the derived magnitude was used to calculate the zero-point. The magnitudes were measured using SExtractor with an aperture radius of 1.6 $\times$ seeing FWHM. The standard deviation of the zero-points from the stars in the field was taken as the zero-point error, found to be $\lesssim 0.05$ magnitudes.

\indent We used the zero-points to derive photometry of the quasar candidates. The same aperture was used for the standard star, and thus the aperture correction was taken into account inherently. The photometry error from SExtractor and the zero-point errors were summed in quadrature. The derived magnitudes were then corrected for the Galactic extinction by applying the extinction law of Cardelli et al. (1989), the dust map of SFD, and assuming $R_{V} = 3.1$.

\section{High-redshift quasar selection}
The selection of quasar candidates was done in two steps, first using the broadband color selection, and then refining the broadband selected sample using the BIC method. This section describes each step of the quasar selection. 

\begin{figure*}[tbh]
\centering 
\includegraphics[scale=0.45]{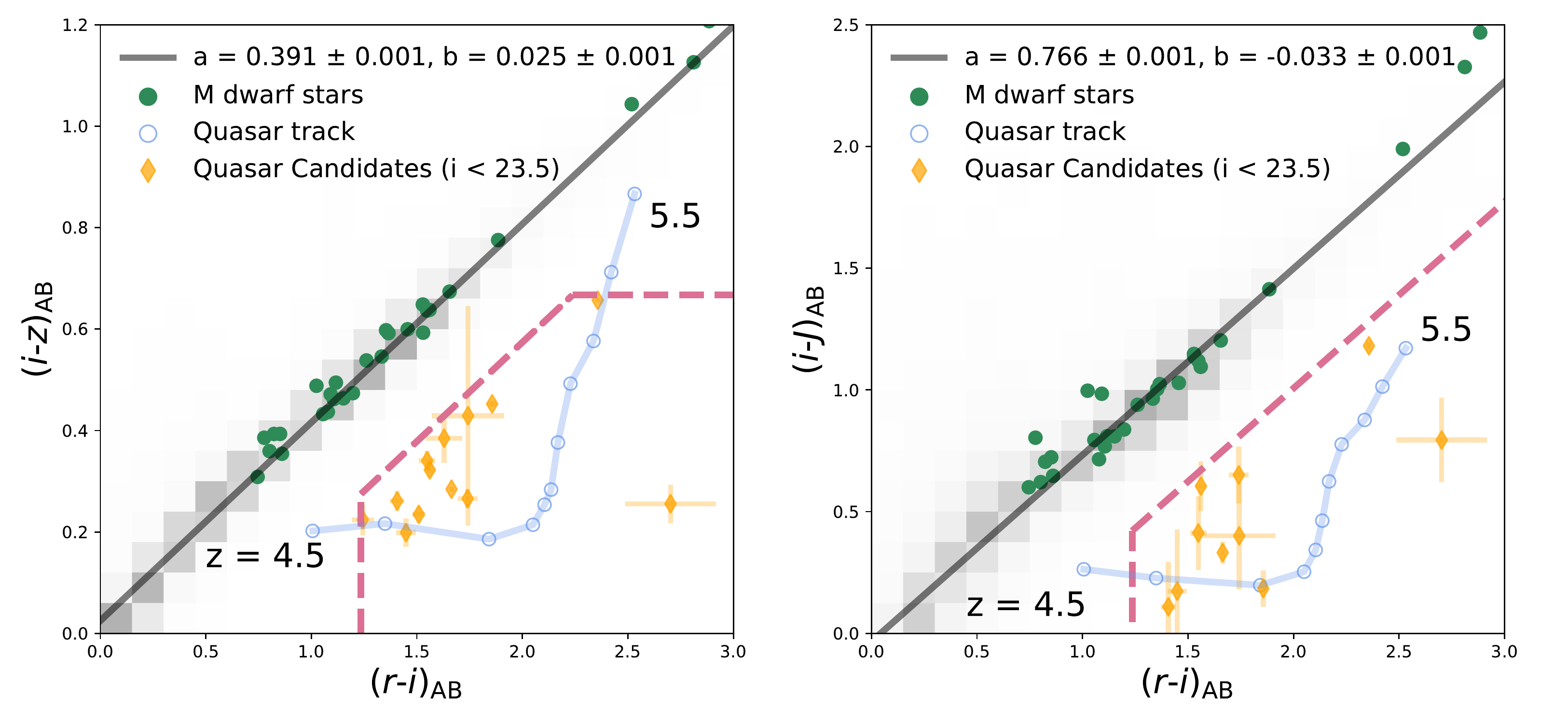}
\caption{Color-color diagrams of point sources at $i < 23.5$ mag. The black shades show the number density of point sources. The quasar selection boxes are marked with the pink dashed line, and the selected quasar candidates are marked with orange diamonds. The blue line indicates the track of a quasar at different redshifts with intervals of 0.1. The green points show the colors of spectroscopically confirmed M dwarf stars from the Fourteenth Data Release of the SDSS (Abolfathi et al. 2018). The M dwarf points follow the stellar locus (the black solid line). On both panels, a and b values indicate the slope and intercept of the linear relation fit to the point sources.}
\label{fig:ccd}
\end{figure*}

\begin{figure}[tbh]
\centering
\includegraphics[scale=0.5]{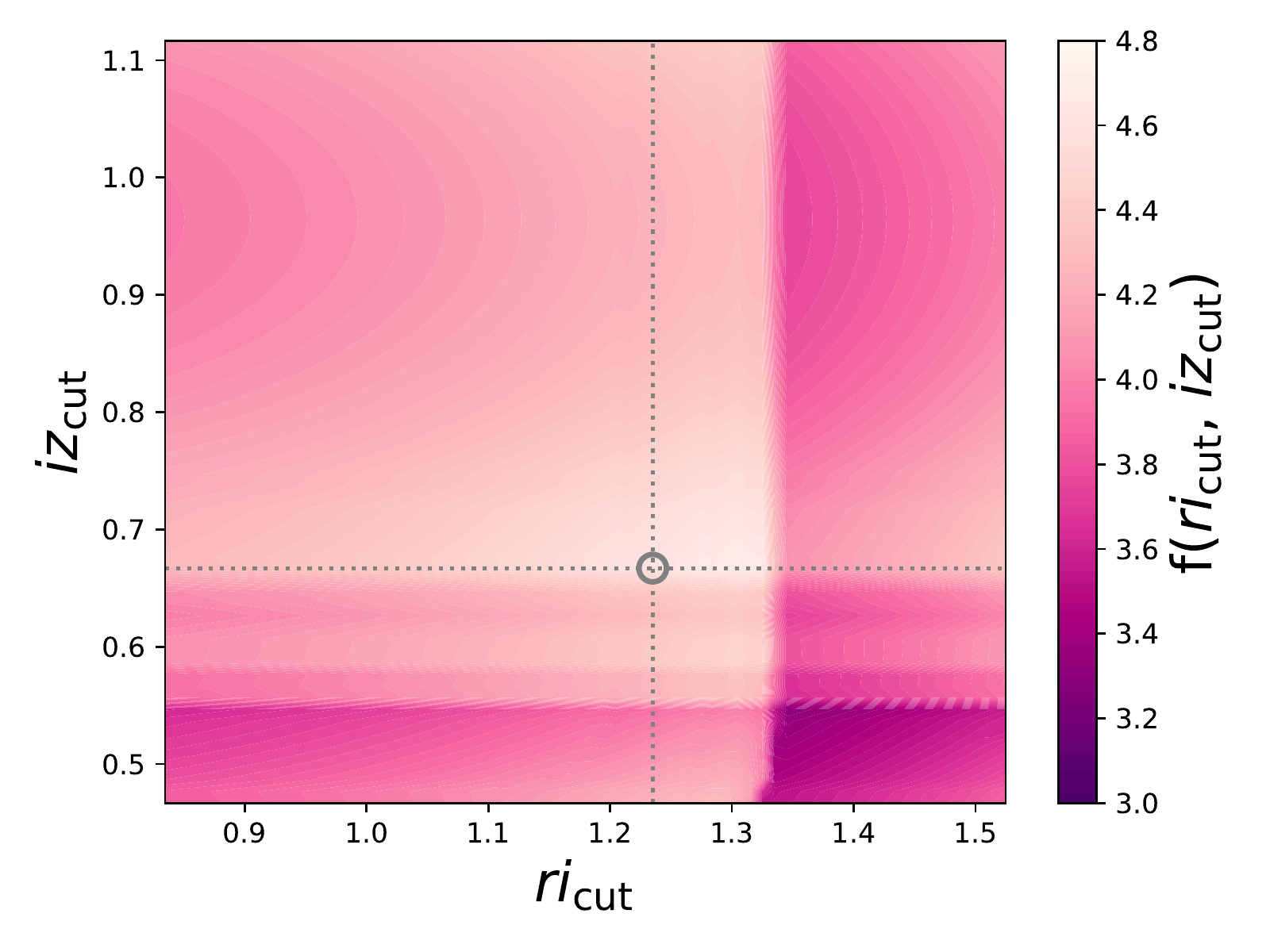}
\caption{The ratio of the completeness to the area of the color selection region (Equation. (1)) as a function of $ri_{\rm{cut}}$ and $iz_{\rm{cut}}$. The dotted line indicates the maximum of the ratio where we define the optimal $ri_{\rm{cut}}$ and $iz_{\rm{cut}}$ values, taking into consideration that one or more quasar model SEDs could be selected with a lower redshift cut at $z=4.5$.}
\label{fig:opt_colorcut}
\end{figure}

\subsection{Broadband color selection criteria}
\indent The quasar candidates were first selected using broadband colors. We employed broadband color cuts in $riz$ and $riJ$ color-color diagrams (CCD), similarly to that described in McGreer et al. (2013) and Kim et al. (2019), with the color-cuts adjusted for the HSC magnitude system.

\indent First, we selected point sources by using the difference between CModel magnitude ($m_{\rm {CModel}}$) and point-spread function (PSF) magnitudes ($m_{\rm{psf}}$) (Matsuoka et al. 2018). Then, we applied $i$ band magnitude cut brighter than 23.5 considering NIR depths. $g$-band flux must be fainter or non-detected due to the IGM attenuation. Thus, we set $g-r > 1.8$ or signal-to-noise ratio (S/N) in $g$-band $< 3.0$ to select objects with weak or non-detection in the $g$ band.

\indent As shown in Figure~\ref{fig:ccd}, the stellar loci can be described with a straight line with a dispersion. Therefore, as a next step, we defined slopes of the stellar loci on both $riz$ and $riJ$ CCDs by fitting a linear line. After getting the slopes of these loci, the distributions of point sources were examined as a function of the distance diagonal from the line. We took percentile cuts of 99.5\% ($riz$) and 95\% ($riJ$) away from the stellar loci to exclude stars from the selection. These are the diagonal pink-dashed lines in Figure~\ref{fig:ccd}.

\indent To further reduce objects that could contaminate the quasar candidate selection, we introduced $r-i$ color cut ($ri_{\rm{cut}}$) and $i-z$ color cut ($iz_{\rm{cut}}$). We defined the $ri_{\rm{cut}}$ and $iz_{\rm{cut}}$ by choosing the condition that include the maximal number of quasar models while reducing the total $riz$ color-space area of the selection box. This was done by choosing the conditions where the following number, $f(ri_{\rm{cut}}, iz_{\rm{cut}})$, is maximal,
\begin{equation}
\setcounter{equation}{1}
f(ri_{\rm{cut}}, iz_{\rm{cut}}) = R(ri_{\rm{cut}}, iz_{\rm{cut}})/ A(ri_{\rm{cut}}, iz_{\rm{cut}}). \\
\end{equation}{}
Here, $R(ri_{\rm{cut}}, iz_{\rm{cut}})$ is the completeness of the quasar selection as defined as the ratio of model quasar SEDs picked up by the selection region to the total number of model quasar SEDs, and $A(ri_{\rm{cut}}, iz_{\rm{cut}})$ is the area defined by the selection region with additional color limits of $ri_{\rm{cut}} < 3$ and $iz_{\rm{cut}} > 0$. The smaller the $A$ value is, the less likely we get contaminants. Also, we optimized the selection criteria by satisfying the prerequisite that one or more quasar model SEDs could be selected with a lower redshift limit at $z=4.5$. 

\indent Figure~\ref{fig:opt_colorcut} shows $f(ri_{\rm{cut}}, iz_{\rm{cut}})$ for various parameters, and we adopted $ri_{\rm{cut}}$ = 1.235 and $iz_{\rm{cut}}$ = 0.673 that maximize $f(ri_{\rm{cut}}, iz_{\rm{cut}})$\footnote{Note $ri_{\rm{cut}} = 1.2$ and $iz_{\rm{cut}} = 0.55$ in McGreer et al. (2013).}. Below, we summarize our broadband color selection criteria:
\begin{enumerate}
\setcounter{enumi}{0}
\item Point sources with $i_{\rm psf}$ - $i_{\rm CModel}$ $<$ 0.15
\item $ i < 23.5 $
\item $g-r > 1.8$ or $S/N(g) < 3.0$ 
\item $r-i > 1.235$ 
\item $i-z < 0.391 (r-i) - 0.220$
\item $i-z < 0.673$ 
\item $i-J < 0.766 (r-i) - 0.525$.
\end{enumerate} 

\begin{deluxetable*}{cccccccccc}[t!]
\tablenum{3}
\tablecaption{SED fitting results} 
\tablehead{\colhead{ID} & \colhead{$z_{\rm{phot}}$} & \colhead{$z_{\rm{spec}}$} & \colhead{$\Delta z / (1 + z_{\rm{spec}})$} & \colhead{$M_{\rm 1450, phot}$} & \colhead{$M_{\rm 1450, spec}$} & \colhead{$\chi^2_{red, d}$} & \colhead{$\chi^2_{red, q}$} & \colhead{$\Delta$BIC}& \colhead{Class}}
\startdata
IMS J160306+541928 & ${4.92}_{-0.05}^{+0.03}$ &  &  & ${-23.42}_{-0.18}^{+0.18}$ &  & 9.2 & 2.9 & 24.7 & quasar \\
IMS J160517+554002 & ${5.16}_{-0.03}^{+0.01}$ & ${5.211}_{-0.001}^{+0.001}$ & 0.008 & ${-24.07}_{-0.14}^{+0.19}$ & ${-24.454}_{-0.017}^{+0.017}$ & 15.3 & 4.1 & 43.0 & quasar \\ 
IMS J160552+555340 & ${5.17}_{-0.03}^{+0.06}$ & ${5.409}_{-0.001}^{+0.001}$ & 0.037 & ${-25.31}_{-0.19}^{+0.21}$ & ${-25.880}_{-0.010}^{+0.010}$ & 18.3 & 2.7 & 46.1 & quasar \\ 
IMS J160622+540056 & ${5.19}_{-0.05}^{+0.03}$ &  &  & ${-23.00}_{-0.20}^{+0.20}$ &  & 12.9 & 4.9 & 13.3 & quasar \\ 
IMS J160732+544750 & ${4.87}_{-0.03}^{+0.06}$ &  &  & ${-22.92}_{-0.08}^{+0.18}$ &  & 24.2 & 8.0 & 58.9 & quasar \\ 
IMS J160748+541157 & ${4.71}_{-0.04}^{+0.09}$ &  &  & ${-23.37}_{-0.13}^{+0.17}$ &  & 12.5 & 2.8 & 38.3 & quasar \\ 
IMS J160914+554511 & ${4.72}_{-0.04}^{+0.04}$ & ${4.814}_{-0.002}^{+0.002}$ & 0.016 & ${-23.61}_{-0.13}^{+0.11}$ & ${-23.807}_{-0.017}^{+0.017}$ & 15.5 & 2.7 & 63.9 & quasar \\ 
IMS J161248+550927 & ${4.67}_{-0.05}^{+0.05}$ &  &  & ${-23.42}_{-0.18}^{+0.16}$ &  & 8.2 & 2.0 & 31.2 & quasar \\ 
IMS J161341+542146 & ${4.68}_{-0.00}^{+0.01}$ &  &  & ${-22.76}_{-0.14}^{+0.16}$ &  & 5.7 & 3.0 & 4.4 & nonquasar \\ 
IMS J161343+542131 & ${5.08}_{-0.04}^{+0.04}$ &  &  & ${-26.08}_{-0.22}^{+0.28}$ &  & 21.4 & 30.0 & -70.8 & nonquasar \\ 
IMS J161636+535545 & ${5.35}_{-0.07}^{+0.01}$ &  &  & ${-23.67}_{-0.13}^{+0.17}$ &  & 12.0 & 4.9 & 11.4 & quasar \\
IMS J161827+551748 & ${4.73}_{-0.01}^{+0.09}$ &  &  & ${-25.12}_{-0.12}^{+0.22}$ &  & 13.4 & 3.4 & 59.1 & quasar \\ 
IMS J161903+545638 & ${4.71}_{-0.02}^{+0.09}$ &  &  & ${-23.99}_{-0.15}^{+0.22}$ &  & 4.1 & 6.1 & -11.8 & nonquasar \\ 
\enddata
\tablecomments{This table show the result of SED-fitting. The errors of $z_{\rm phot}$ and $M_{\rm {1450, phot}}$ are derived from the 68\% confidence intervals with other parameters fixed, while $z_{\rm spec}$ and $M_{\rm {1450, spec}}$ are estimated from MCMC simulation with other parameters free. The BIC values are calculated using all photometric data.}
\label{table:BIC}
\end{deluxetable*}

\indent Quasar candidates that satisfy the color selection were visually inspected and rejected if the photometry is found to be spurious during the visual inspection (e.g., scattered light contamination). Only one source was found to be spurious and rejected during the visual inspection. Finally, we identified 13 quasar candidates, which are shown in Figure~\ref{fig:ccd}. Table~\ref{table:grizJmag} summarizes the photometry of the quasar candidates.

\subsection{Bayesian information criterion (BIC) selection and SED fitting}
\subsubsection{SED fitting}
To refine the quasar selection, we first fitted the SED of the quasar candidates to various quasar and M-dwarf models. For the SED fitting, we added the medium-band data to sample the sharp break in the quasar SED located at the wavelength of redshifted Ly$\alpha$. We then used the information from the fitting to apply the BIC to refine the candidate selection.

\begin{figure*}[tbh]
\includegraphics[width=0.995\textwidth]{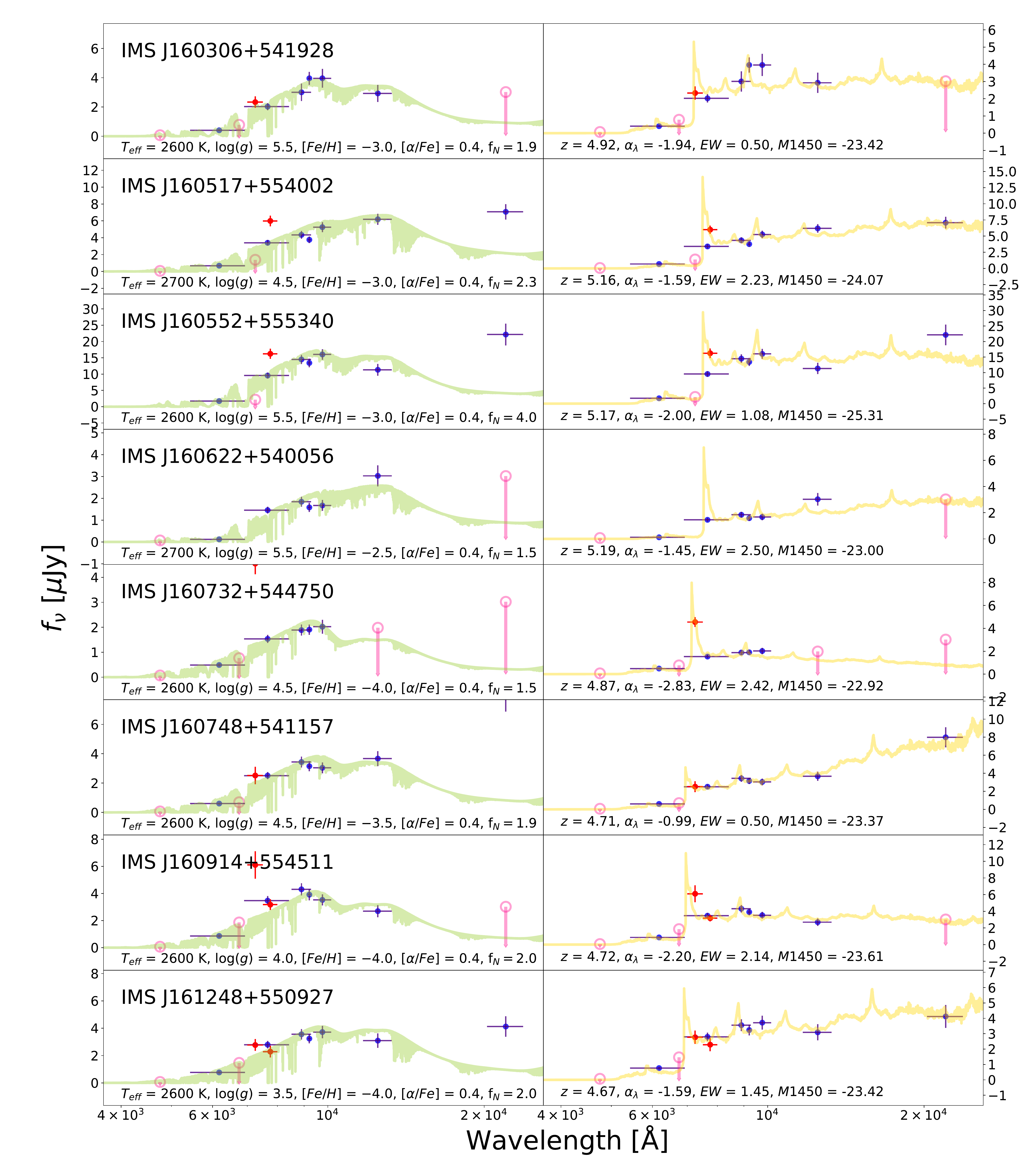} 
\caption{SED-fitting results of the broadband selected quasar candidates. The broadband data are expressed as blue points, while the medium-band data are plotted with red points. The horizontal bar of each point indicates the width of the filter. For non-detections, pink circles and arrows present the 5-$\sigma$ limiting magnitudes. The best dwarf star model and the best quasar model obtained through the SED fitting are shown as the green and gold lines, respectively. The best-fit results are also indicated, where f$_{N}$ are in units of $10^{-12}$.}
\label{fig:SEDfitting}
\end{figure*}

\indent The SED fitting provided the best-fit parameters and the goodness-of-fit in terms of the reduced chi-squares, $\chi_{\rm red}^{2}$, for each model. For quasar models, we obtained the photometric redshift $z_{\rm{phot}}$, the continuum slope $\alpha_{\lambda}$, the EW of the Ly$\alpha$+N\Romannum{5} line, and the UV absolute magnitude at 1450 $\AA$, $M_{\rm 1450}$. For M-dwarf models, we obtained $T_{\text{eff}}$, log($g$), [M/H], [$\alpha$/M], and the normalization parameter. The SED-fitting followed the procedure as described in Kim et al. (2019), and it takes into account the upper limits as described in by Sawicki (2012). The results of the SED fitting are shown in Figure~\ref{fig:SEDfitting}. In Table~\ref{table:BIC}, we present the derived parameters such as $z_{\rm{phot}}$, $M_{\rm 1450}$, and $\chi_{\rm red}^{2}$ for both the best-fit quasar and M-dwarf models. Figure~\ref{fig:SEDfitting} shows that many of the candidates are better fitted to the quasar models (e.g., IMS J160914+554511), but some are not (e.g., IMS J161343+542131) and some are ambiguous (e.g., IMS J161341+542146). In order to better differentiate quasars from M-dwarfs, we apply the BIC as described in the next subsection.

\renewcommand{\thefigure}{\arabic{figure} (Cont.)}
\addtocounter{figure}{-1}
\begin{figure*}[tbh]
\includegraphics[width=0.995\textwidth]{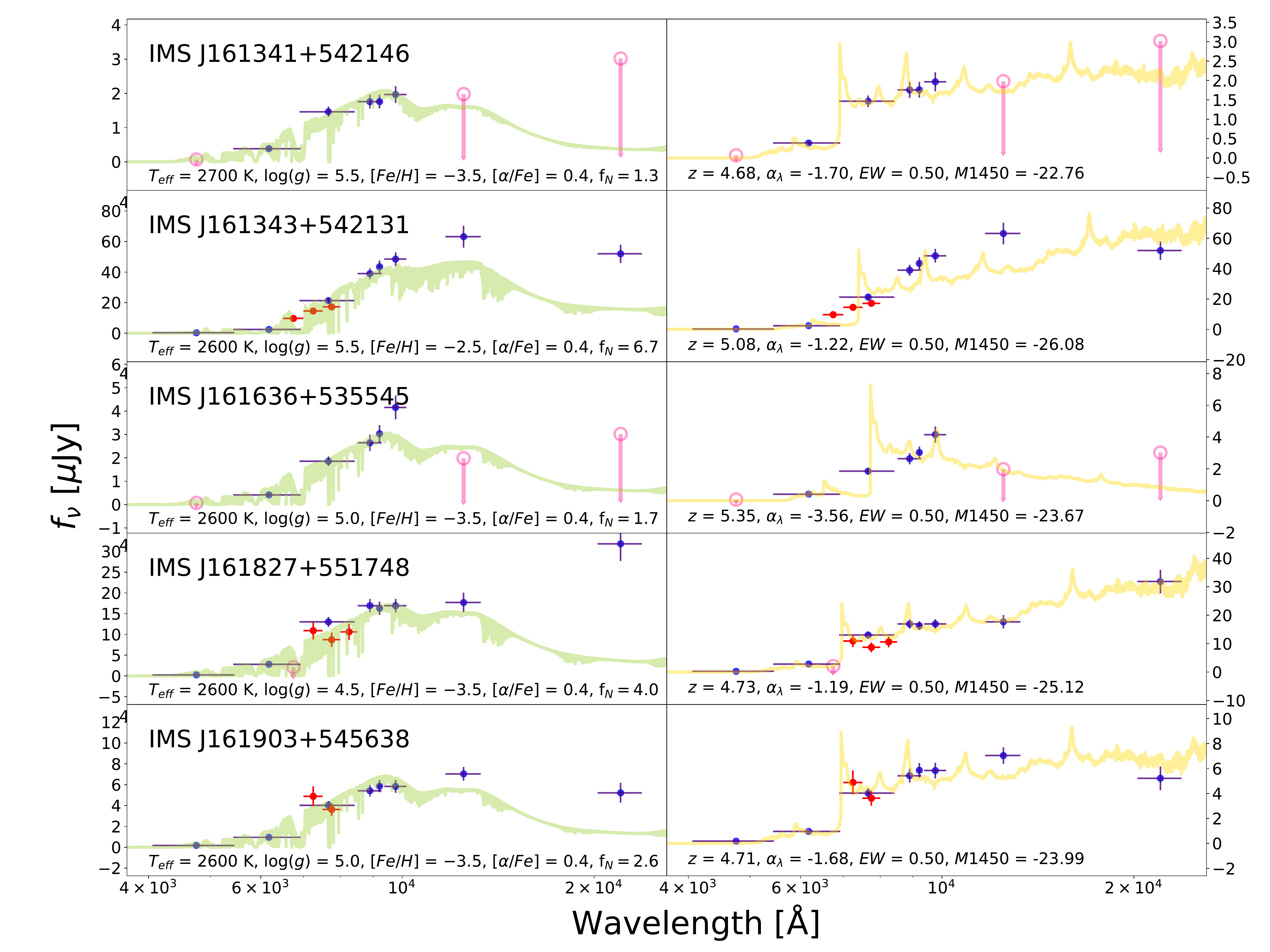} 
\caption{}
\end{figure*}
\renewcommand{\thefigure}{\arabic{figure}}

\begin{figure*}[tbh]
\centering 
\includegraphics[width=0.5\textwidth]{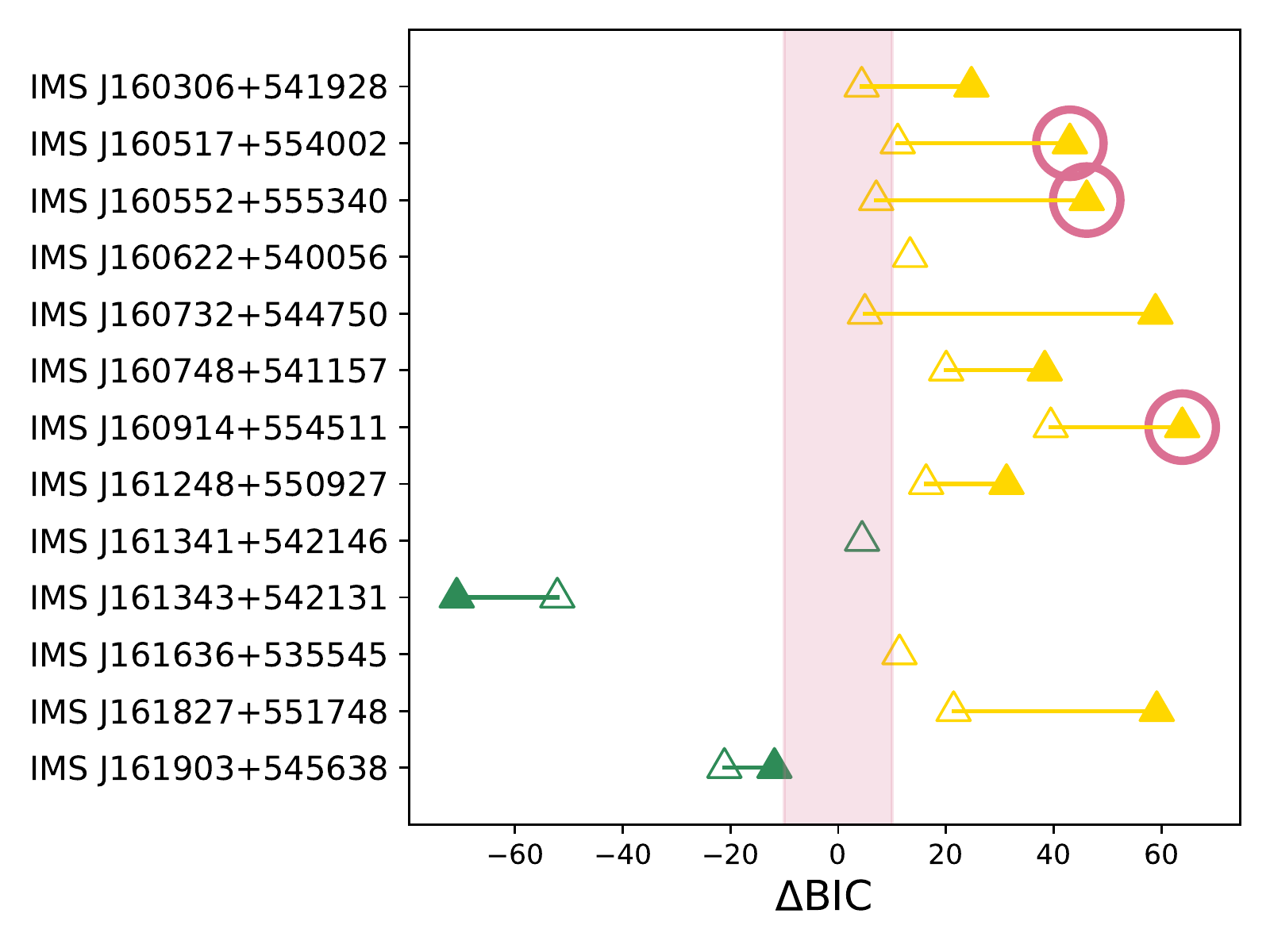}
\caption{$\Delta$BIC graph showing if a broadband selected quasar is more like stars or quasars. The pink shaded area indicates the region with $\abs{\Delta \rm BIC} < 10$ where it is difficult to choose the better model with confidence. The quasars should lie on the right side of the shaded area, while M-dwarf should lie on the left. The filled and open triangles are $\Delta$BIC of the candidates with and without medium-band data, respectively. The spectroscopically confirmed quasars are marked with the open pink circles. The addition of medium-band data clearly helps distinguish which objects are quasars, with the same classification uncertain otherwise.}
\label{fig:BIC}
\end{figure*}

\begin{figure*}[tbh!]
\centering
\includegraphics[width=0.75\textwidth]{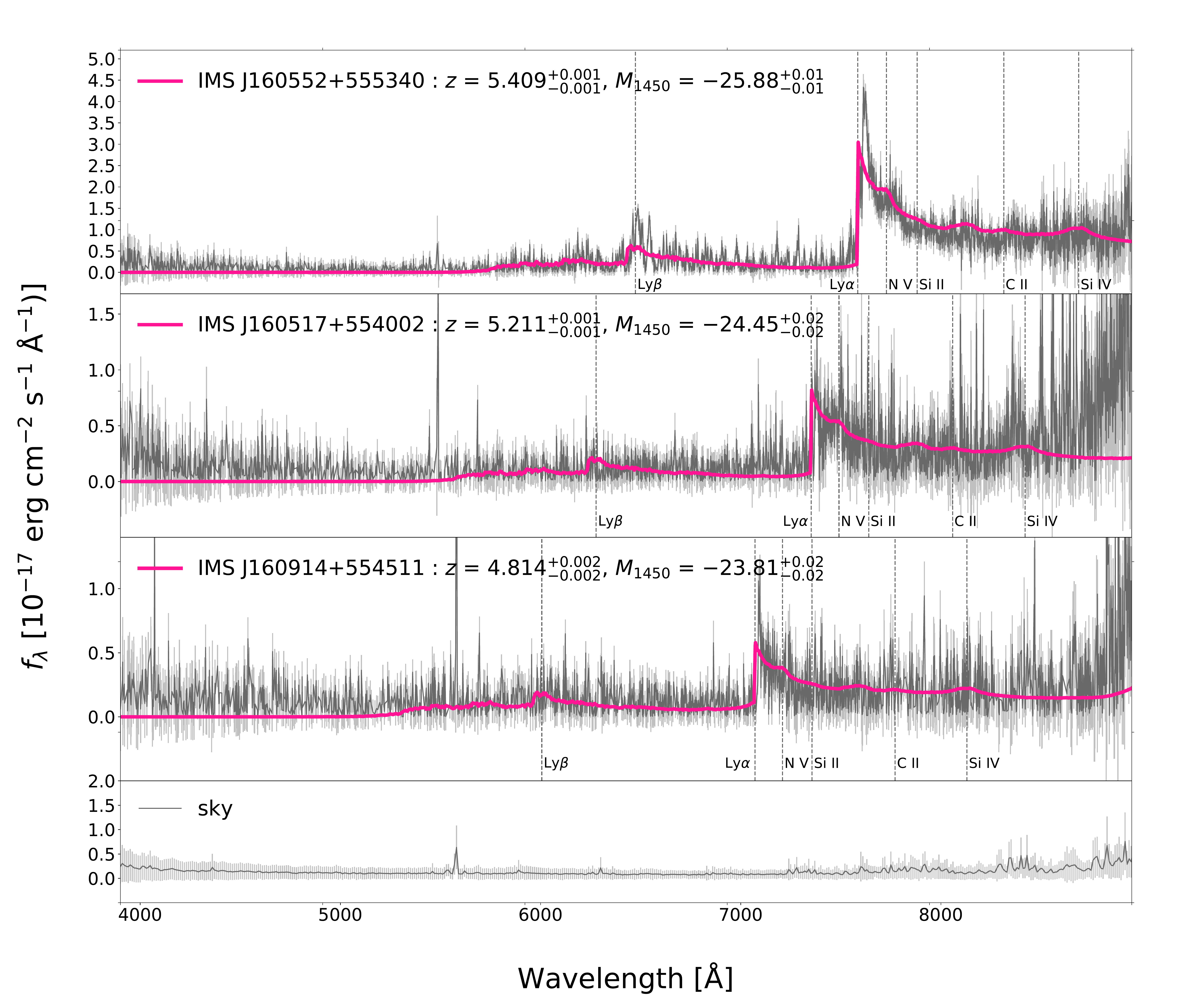}
\caption{MMT Hectospec spectra of the three spectroscopically confirmed quasars (black lines) with 1$\sigma$ flux uncertainties (gray lines). The pink lines show the best-fit results of quasar model SED-fitting. The sky signal and its errors are presented in the bottom panel. The sharp breaks at the redshifted Ly$\alpha$ indicate that they are all high redshift quasars. The $z_{\rm{spec}}$ and the $M_{\rm 1450}$ values from the SED fitting are shown in the figure.}
\label{fig:Spectra}
\end{figure*}

\subsubsection{BIC selection of quasars}
To improve the quasar selection, we used the BIC for each best-fit model rather than $\chi_{\rm red}^2$. The ratio of the $\chi_{\rm red}^{2}$ of best-fit quasar model to $\chi_{\rm red}^{2}$ of best-fit M-dwarf star model can be used to prioritize the candidates (Mazzucchelli et al. 2017). However, the ratio cannot consider the effect of difference in the number of fitting parameters between two models. On the contrary, the BIC can take the number of free parameters as well as the number of data into account, which can be defined as 
\begin{equation}
\setcounter{equation}{2}
\text{BIC} = -2 \ln L_{\rm{max}} + k \ln n \\,
\end{equation}
where $n$ is the number of photometric data, $k$ is the number of free parameters in the tested model, and $L_{\rm{max}}$ is the maximum likelihood value of the model. Therefore, the BIC can be an effective tool when comparing models with the different $k$.

\indent The difference in the values of BIC of two models can be a quantitative measure to tell between the better-fitting model and given as, 
\begin{equation}
\setcounter{equation}{3}
\Delta\text{BIC} = [\chi^2 + k \ln n ]_{d} - [\chi^2 + k \ln n]_{q},
\end{equation}
where `$d$' and `$q$' in subscripts indicate the best-fitting M-dwarf star and quasar models, respectively. If the difference larger than 10 in BIC commonly means `decisive' evidence that supports the quasar model is better than the M-dwarf star model (Liddle 2007). 

\indent We computed the $\Delta$BIC of the best-fit quasar and the M-dwarf models with and without medium-band data (see Figure~\ref{fig:BIC} and Table~\ref{table:BIC}). Among 13 candidates from the broadband color selection, we identified 10 to be promising high-redshift quasar candidates, and 2 to be M-dwarf stars. The medium-band data are critical for distinguishing the nature of the candidates, as one of the three spectroscopically identified quasars in Section 5 are deemed ambiguous unless the additional observation data were used.

\section{Spectroscopic confirmation of quasars}
\label{sec:spectro}
\subsection{Observation}
Three quasar candidates were observed on 2018 July 10 with the Hectospec instrument on MMT (Fabricant et al. 2005). These candidates were chosen based on the $\Delta$BIC values in Table~\ref{table:BIC} and the feasibility of simultaneous observations with other objects. The Hectospec can place fibers on $\sim$ 300 targets within a one-degree diameter field of view, and the quasar targets were included as a part of another program studying galaxy clusters in the ELAIS-N1 field. For the observation, we used the 270 lines$/$mm grating that covers the wavelength range of 3650--9200 $\rm{\AA}$ at a spectral resolution of $\sim1000$. On-source exposure time was 240 minutes per quasar. The spectra were reduced with the HSRED pipeline for the basic reduction (bias and flat-field), the wavelength and flux calibrations, and the sky subtraction. The standard star and the sky fiber data were used for the flux calibration and the sky subtraction, respectively.

\begin{figure}[tbh]
\centering 
\includegraphics[width=0.5\textwidth]{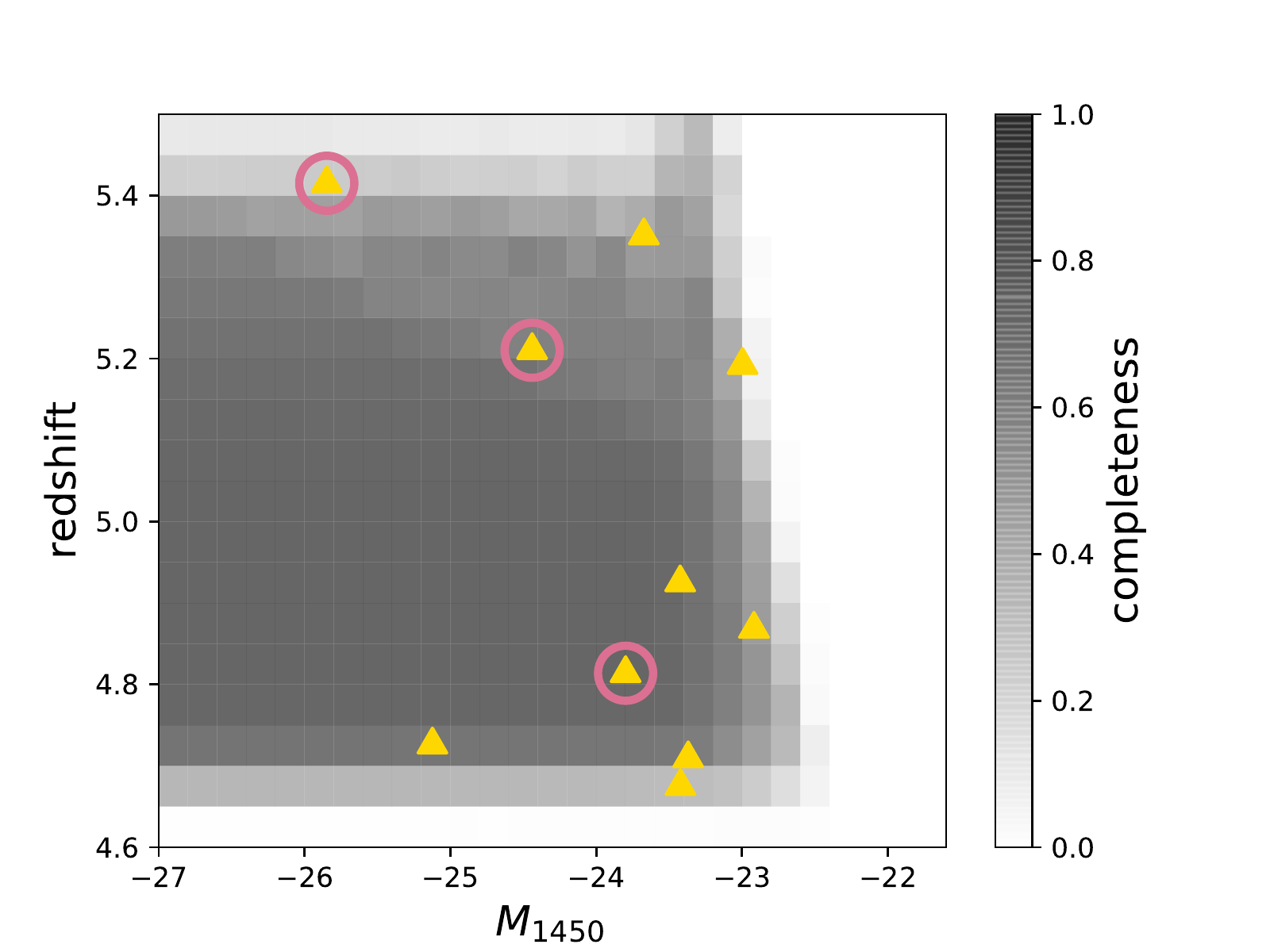}
\caption{The completeness function ($F_{comp}$) of the broadband quasar selection as a function of $z$ versus $M_{\rm 1450}$. This figure indicates that quasars at z = $4.6\lesssim z \lesssim 5.4$ in the magnitude range of $-27 \lesssim M_{\rm 1450} \lesssim -23 $ will be efficiently selected. The gold triangle shows BIC-selected sample. The spectroscopically confirmed quasars in the sample are marked with the open pink circles.}
\label{fig:Completeness}
\end{figure}

\subsection{Spectroscopic identification}
The Hectospec observation reveals all the candidates, IMS J160517+554002, IMS J160552+555340, and IMS J160914+554511, are quasars at $z\sim5$. Figure~\ref{fig:Spectra} shows their spectra. The strong break at Ly$\alpha$ can be seen in all spectra, suggesting that they are all indeed at $z\sim5$. Their spectroscopic redshifts and $M_{\rm 1450}$ are derived with the SED-fitting using 20,000 chains of Markov Chain Monte Carlo (MCMC) simulation to these spectra. Both $z_{\rm{spec}}$ and $z_{\rm{spec}}$-based $M_{\rm 1450}$ of the three quasars are given in Table~\ref{table:BIC}. Their $M_{\rm 1450}$ have ranges $\ga$ -26. The 3/3 confirmation rate gives us the confidence for the BIC selection of $z\sim5$ quasars including medium-band observations.

\section{Luminosity Function and\newline implication for reionization}
\begin{figure*}[tbh!]
\centering 
\includegraphics[scale=0.5]{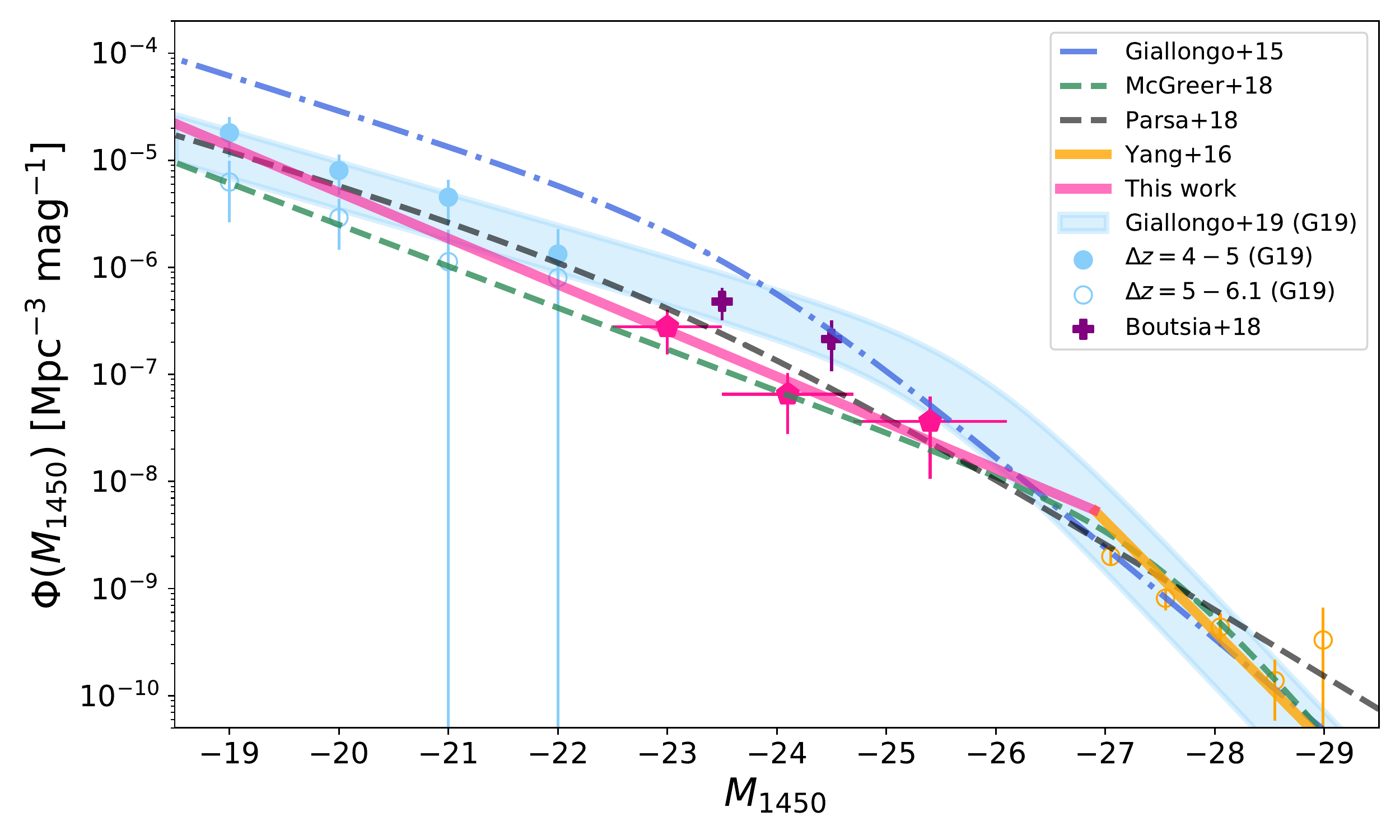}
\caption{The quasar luminosity function at $z\sim5$ from this work (pink pentagons and solid line). The x-axis and y-axis error bars of the number density from this work represent the luminosity bin width for $V_{a}$ calculation and the statistical error of the number density estimate, respectively. The orange points indicate the quasar LF from Yang et al. (2016). Also plotted are the fitted LF functions from G15 (the blue dotdashed line), McGreer et al. (2018; the green dashed line), Parsa et al. (2018; the black dashed line) and G19 (sky-blue line with shade). Note that LFs of G19 at $z\sim4.5$ and $5.6$ are shown as the filled and open sky-blue circles, respectively, and we show both functions with a shade. The purple plus markers are number densities of Boutsia et al. (2018) rescaled to $z=5$.}
\label{fig:QLF}
\end{figure*}

From the ten BIC-selected candidates at a redshift range 4.6 $\lesssim$ z $\lesssim$ 5.4 including three confirmed quasars, we derive the quasar luminosity function at $M_{\rm 1450} \ga -26$ mag. To do so, we use the updated 1/$V_{a}$ method (Page \& Carrera. 2000; Im et al. 2002). For a given interval $\Delta z$ and a given magnitude interval $\Delta M_{\rm 1450}$, $V_{a}$, the effective volume covered by $N$ number of quasars belonging in the bin can be written as, 
\begin{equation}
\setcounter{equation}{4}
V_{a}\,\Delta M_{\rm 1450} = \int_{\Delta M_{1450}} \int_{z_{\rm min}}^{z_{\rm max}(M_{1450})} \! F_{\rm comp}\,\frac{\mathrm{d}V}{\mathrm{d}z}\,\mathrm{d}z\,\mathrm{d} M_{1450}\,,\\
\end{equation}
where $F_{comp}$ represents the completeness from 0.0 to 1.0, ${\mathrm{d}V}/{\mathrm{d}z}$ is the cosmological volume element taking into account of the survey area, $z_{min}$ is the lowest redshift of the redshift interval, and $z_{max}(M_{\rm 1450})$ is the maximum redshift where an object with $M_{\rm 1450}$ is within the flux limit of the survey.

\indent The completeness function $F_{\rm comp}$ is the fraction of quasar models enable to pass the selection criteria respect to all the simulated quasar models at given ranges of parameters. The completeness function has two variables of redshift and magnitude, which are binned with a bin size of 0.05 and 0.2, respectively. Figure~\ref{fig:Completeness} show the completeness function values across $4.5 < z< 5.5$ and $-27 < M_{\rm 1450} < -23$. It supports the relevance of our survey searching for quasar candidates with $z\sim 5$ and $M_{\rm 1450} \la -23$.

\indent Note that we assumed that the photometric completeness is 100$\%$ considering that the point source detection completeness of a similar depth HSC data is 100 $\%$ at $i<23.5$ (Matsuoka et al. 2018). Similarly, we did not take into account, in the $F_{\rm comp}$ calculation, the point source completeness and the incompleteness in the sample due to source confusion. The point source completeness of a bit shallower HSC data is $\gtrsim 95$ \%, and the source confusion affects the $F_{\rm comp}$ at a similar level (Matsuoka et al. 2018). They are much smaller than the errors of the LF points which amounts to 40 \% or larger (Table 4), and thus are neglected in the LF calculation.

\indent Then, we calculate the binned number density and its uncertainty, $\delta N$ according to the following equations where $N$ is the number of the observed objects in the magnitude bin, and $\delta \Phi$ is directly related to the Poisson noise of $N$.
\begin{equation}
\setcounter{equation}{5}
\Phi = \frac{N}{V_{a}\,\Delta M_{\rm 1450}}, \qquad \delta \Phi = \frac{\sqrt{N}}{V_{a}\,\Delta M_{\rm 1450}}\\
\end{equation}
\indent Due to the small number of the sample, we adopt a simple power-law function for the luminosity function as $\Phi = \Phi_{*}\,L^{\alpha+1}$, or as, 
\begin{equation}
\setcounter{equation}{6}
\text{log}_{10}\,(\,\Phi_{\rm model}(M_{\rm 1450})\,) = a\;M_{\rm 1450} + b. \\
\end{equation}
Here, $a$ is related to the slope of the luminosity function $\alpha$, as $\alpha=-2.5 \times a -1$, and $b$ is related to the normalization of $\Phi$. We estimate the parameters in $\Phi_{model}$ by minimizing $S = -2 \ln{L}$, described as,  
\begin{multline}
\setcounter{equation}{7}
S = -2 \sum \ln [\Phi_{\rm model}(M_{1450})F_{\rm comp}(z, M_{\rm 1450})] \\ 
+ 2 \int \int \Phi_{\rm model}(M_{1450})F_{\rm comp}(z,M_{1450})\frac{\mathrm{d}V}{\mathrm{d}z}\,\mathrm{d}z\,\mathrm{d} M_{1450}. 
\end{multline}

\noindent The first term in $S$ is applicable to the newly discovered quasars and the BIC-selected candidates. The second term in $S$ provides the normalization term, and summed over the whole redshift and the magnitude ranges of the sample. Note that, in this parameter estimation, we ignore the redshift evolution in the quasar number density, since the number density evolution is negligible over this redshift range of 4.6 $\lesssim$ z $\lesssim$ 5.4.

\begin{deluxetable}{cccccc}[tbh] 
\tablenum{4}
\label{table:Shinetal}
\tablecaption{Quasar LF} 
\tablehead{\colhead{${M_{\rm 1450}}$} & \colhead{$\Delta{M_{\rm 1450}}$} & \colhead{$\Phi$} & \colhead{$\sigma_{\Phi}$} & \colhead{$N$} &\colhead{$N_{\rm{cor}}$} \\ \colhead{} & \colhead{} & \colhead{Mpc$^{-3}$ mag$^{-1}$} & \colhead{} & \colhead{} & \colhead{} }
\startdata
-25.4 & 1.4 & 3.64e-08 & 2.57e-08 & 2.0 & 4.3 \\
-24.1 & 1.2 & 6.52e-08 & 3.78e-08 & 3.0 & 3.9 \\
-23.0 & 1.0 & 2.79e-07 & 1.25e-07 & 5.0 & 18.6 \\
\hline
& & & & & \\
\hline
\hline
& $a$ & & & $b$ & \\
\hline
& $0.43_{-0.26}^{+0.29}$ & & & $3.3_{-6.3}^{+6.9}$ & \\
\enddata
\end{deluxetable}

\indent The binned LF and the fitted parameter values are presented in Table~\ref{table:Shinetal}. Also, Figure~\ref{fig:QLF} shows the binned LF and the fitted LF, along with several recent LF values from the literature. The fitted LF slope at $M_{\rm 1450} \gtrsim -26$ is ${-2.08}_{-0.72}^{+0.65}$, which is in line with recently reported faint end slope of ${-1.97}_{-0.09}^{+0.09}$ (M18), but is slightly steeper than the results from some other studies (Akiyama et al. 2018; Matsuoka et al. 2018; G19). 

\indent More importantly, the LF at $-26 \lesssim M_{\rm 1450} \lesssim -23$ of our study gives a much lower density of quasars ($\sim6$ times) at $z\sim5$ than the earlier study of G15. Our LF is in line with that of M18, and is not too far from the more recent results from Boutsia et al. (2018) rescaled to $z=5$ and G19. To estimate the UV emissivity contribution of the quasars within the $M_{1450}$ range of -29 to -18, we adopt the bright-end slope of the best-fit quasar LF in Yang et al. (2016) and expressing the LF with single power-law function. We then combine our LF with the LF of Yang et al. (2016) and merge them at the intersect of the two LFs. By assuming the escape fraction of 1 (Cristiani et al. 2016; Guaitaet al. 2016; Grazian et al. 2018; Romano et al. 2019) and following the equation (3) in G15, we find $\epsilon_{912}\footnote{The LyC emissivity $\epsilon_{912}$ is in units of $10^{24}$ erg s$^{-1}$ Hz$^{-1}$ Mpc$^{-3}$.} \sim 1$. Comparing to $\epsilon_{912} \sim 0.8$ of McGreer et al. (2018), $\epsilon_{912} \sim 1.2$ of Parsa et al. (2018) and $\epsilon_{912} \sim 3.8$ of G19, our result prefers a minor contribution of a quasar to keep the IGM ionized at $z\sim5$. The photoionization rate from our result is $\sim0.03 \times 10^{-12}$ s$^{-1}$, which is $\gtrsim10$ times smaller than the photoionization rate of UV background at $z\simeq5$ (Bolton et al. 2007, Calverley 2011, Wyithe \& Bolton 2011). Therefore, optically selected quasars are insufficient to fully explain the IGM ionization at $z\sim5$.

\section{Summary}
In this study, we searched for faint quasars at $z\sim5$ with $-26 \lesssim M_{\rm 1450} \lesssim -23$ ($i<23.5$ mag) in the ELAIS-N1 field over an area of 6.51 deg$^2$. Among optical quasars, quasars in this magnitude range are thought to contribute the most to the IGM ionizing photons, therefore, securing a sample of such quasars is important for understanding the cosmic IGM ionization history.

\indent Using the devised broadband color selection criteria to optimize for the HSC photometry, we identified 13 $z\sim5$ quasar candidates with the optical/NIR imaging data from HSC-SSP, IMS and DXS survey. We then refined the candidate selection by adding medium-band data and performing additional cuts based on BIC. As a result, we obtained the refined sample of 10 candidates. The spectroscopic observation was carried out for three of the BIC-selected candidates using MMT Hectospec, and all of these candidates were identified as quasars at $z\sim5$. In addition, the redshift and $M_{\rm 1450}$ of the other BIC-selected candidates were derived through the SED-fitting.

\indent Using the three newly confirmed quasars and the seven BIC-selected quasar candidates with $z_{\rm{phot}}$, we constructed the $z\sim5$ quasar LF at $-26 \lesssim M_{\rm 1450} \lesssim -23$, with a single power-law function with a slope of ${-2.08}_{-0.72}^{+0.65}$. We also found that the number density of the faint $z\sim5$ quasars is $\sim 10^{-7}$ Mpc$^{-3}$ mag$^{-1}$. Combined with the bright-end LF from the literature, we showed that the UV emissivity of quasars are insufficient to explain the IGM ionization at $z\sim5$, in line with the earlier results for $z\gtrsim 6$ quasars.

\indent Our method of finding high-redshift quasars via medium-band data and BIC can improve the efficiency of quasar survey. The broadband color selection is an efficient way to identify high redshift quasars, but 3 among the 10 BIC-selected candidates (30\%) has been unambiguously selected as quasars from the broadband colors only. These candidates needs the additional medium-band observation to make sure they are highly promising candidates. The medium-band observations can be done with small to mid-sized telescope for which telescope time is more readily available than larger class telescopes, and we expect that future surveys of faint quasars would benefit from medium-band observations when the broadband selection is uncertain.

\acknowledgments
\indent We appreciate the anonymous referee's constructive comments. We thank Yoonsoo Bach for his input on the statistical analysis. M.H. acknowledges the support from Global PH.D Fellowship Program through the National Research Foundation of Korea (NRF) funded by the MInistry of Education (NRF-2013H1A2A1033110)
\indent This work was supported by the National Research Foundation
of Korea (NRF), grant No. 2017R1A3A3001362, No. 2017R1A6A3A04005158 and No. NRF-2019R1C1C1002796, funded by the Korea government (MSIP). This work was supported by the K-GMT Science Program (PID:2018B-UAO-G7) of the Korea Astronomy and Space Science Institute (KASI). 
Based [in part] on data collected at the Subaru Telescope and retrieved from the HSC data archive system, which is operated by Subaru Telescope and Astronomy Data Center at National Astronomical Observatory of Japan. The Hyper Suprime-Cam (HSC) collaboration includes the astronomical communities of Japan and Taiwan, and Princeton University. The HSC instrumentation and software were developed by the National Astronomical Observatory of Japan (NAOJ), the Kavli Institute for the Physics and Mathematics of the Universe (Kavli IPMU), the University of Tokyo, the High Energy Accelerator Research Organization (KEK), the Academia Sinica Institute for Astronomy and Astrophysics in Taiwan (ASIAA), and Princeton University. Funding was contributed by the FIRST program from Japanese Cabinet Office, the Ministry of Education, Culture, Sports, Science and Technology (MEXT), the Japan Society for the Promotion of Science (JSPS), Japan Science and Technology Agency (JST), the Toray Science Foundation, NAOJ, Kavli IPMU, KEK, ASIAA, and Princeton University. This paper makes use of software developed for the Large Synoptic Survey Telescope. We thank the LSST Project for making their code available as free software at  http://dm.lsst.org. The Pan-STARRS1 Surveys (PS1) have been made possible through contributions of the Institute for Astronomy, the University of Hawaii, the Pan-STARRS Project Office, the Max-Planck Society and its participating institutes, the Max Planck Institute for Astronomy, Heidelberg and the Max Planck Institute for Extraterrestrial Physics, Garching, The Johns Hopkins University, Durham University, the University of Edinburgh, Queen’s University Belfast, the Harvard-Smithsonian Center for Astrophysics, the Las Cumbres Observatory Global Telescope Network Incorporated, the National Central University of Taiwan, the Space Telescope Science Institute, the National Aeronautics and Space Administration under Grant No. NNX08AR22G issued through the Planetary Science Division of the NASA Science Mission Directorate, the National Science Foundation under Grant No. AST-1238877, the University of Maryland, and Eotvos Lorand University (ELTE) and the Los Alamos National Laboratory.
The UKIRT is owned by the University of Hawaii (UH) and operated by the UH Institute for Astronomy; operations are enabled through the cooperation of the East Asian Observatory. When (some of) the data reported here were acquired, UKIRT was operated by the Joint Astronomy Centre on behalf of the Science and Technology Facilities Council of the U.K. 
Observations reported here were obtained at the MMT Observatory, a joint facility of the Smithsonian Institution and the University of Arizona. 
This paper includes data taken at the McDonald Observatory of the University of Texas at Austin. 

\indent Facilities: UKIRT (WFCAM), Struve (SQUEAN), MMT (Hectospec), Sloan.

\indent Software: SExtractor (Bertin \& Arnouts 1996), Astrometry.net (Lang et al. 2010)

\end{document}